\documentclass[paper,11pt]{JHEP}
\usepackage[centertags]{amsmath}
\usepackage{amsfonts} \usepackage{amssymb} \usepackage{amsthm}
\allowdisplaybreaks[1]
\usepackage{graphicx}
\newcommand{\Z}{{\mathbb Z}}

\newcommand{\C}{{\mathbb C}}
\newcommand{\vv}{\mathbf{v}}
\newcommand{\qen}{q_{\mathrm{enh}}}
\newcommand{\eq}[1]{Eq.~(\ref{#1})}

\newcommand{\ii}{\mathrm{i}}
\newcommand{\ee}{\mathrm{e}}

\newcommand{\vev}[1]{\langle #1 \rangle}

\newcommand{\tr}{\mathrm{tr}}
\newcommand{\diag}{\mathrm{diag}\,}

\newcommand{\cM}{{\mathcal{M}}}
\newcommand{\cN}{{\mathcal{N}}}
\newcommand{\cO}{{\mathcal{O}}}

\newcommand{\cV}{{\mathcal{V}}}

\newcommand{\one}{{\rm 1\kern -.9mm l}}

\newcommand{\ch}[2]{{\textstyle{\Big[\begin{array}{c} #1\\ #2\end{array}\Big]}}}

\newcommand{\beq}{\begin{equation}}
\newcommand{\eeq}{\end{equation}}
\newcommand{\bea}{\begin{eqnarray}}
\newcommand{\eea}{\end{eqnarray}}
\newcommand{\ba}{\begin{array}}
\newcommand{\ea}{\end{array}}
\newcommand{\ft}[2]{{\textstyle\frac{#1}{#2}}}
\def\nn{\nonumber}

\newdimen\tableauside\tableauside=1.0ex
\newdimen\tableaurule\tableaurule=0.4pt
\newdimen\tableaustep
\def\phantomhrule#1{\hbox{\vbox to0pt{\hrule height\tableaurule
width#1\vss}}}
\def\phantomvrule#1{\vbox{\hbox to0pt{\vrule width\tableaurule
height#1\hss}}}
\def\sqr{\vbox{%
  \phantomhrule\tableaustep
\hbox{\phantomvrule\tableaustep\kern\tableaustep\phantomvrule\tableaustep}%
  \hbox{\vbox{\phantomhrule\tableauside}\kern-\tableaurule}}}
\def\squares#1{\hbox{\count0=#1\noindent\loop\sqr
  \advance\count0 by-1 \ifnum\count0>0\repeat}}
\def\tableau#1{\vcenter{\offinterlineskip
  \tableaustep=\tableauside\advance\tableaustep by-\tableaurule
  \kern\normallineskip\hbox
    {\kern\normallineskip\vbox
      {\gettableau#1 0 }%
     \kern\normallineskip\kern\tableaurule}%
  \kern\normallineskip\kern\tableaurule}}
\def\gettableau#1 {\ifnum#1=0\let\next=\null\else
  \squares{#1}\let\next=\gettableau\fi\next}
\newcommand{\Yfund}{\tableau{1}}
\newcommand{\Ysymm}{\tableau{2}}
\newcommand{\Yasymm}{\tableau{1 1}}

\title{Non-perturbative gauge/gravity correspondence in $\mathcal N=2$ theories}
\author{M. Bill\'o$^1$, M. Frau$^{1}$, F. Fucito$^{2}$, L. Giacone$^{1}$, A. Lerda$^{3}$, J.~F. Morales$^{2}$, D. Ricci Pacifici$^{2}$
\\
\vskip 0.2cm
$^1$ Universit\`a degli Studi di Torino, Dipartimento di Fisica\\
and I.N.F.N. - Sezione di Torino \\
Via P. Giuria 1, I-10125 Torino, Italy\\
\vskip 0.2cm
$^2$ I.N.F.N. - Sezione di Roma 2\\
and Universit\`a di Roma Tor Vergata, Dipartimento di Fisica\\
Via della Ricerca Scientifica, I-00133 Roma, Italy\\
\vskip 0.2cm
$^3$ Universit\`a del Piemonte Orientale, Dipartimento di Scienze e Innovazione Tecnologica\\
and I.N.F.N. - Gruppo Collegato di Alessandria - Sezione di Torino\\
Viale T. Michel  11, I-15121 Alessandria, Italy\\
\vspace{0.25cm}
\email{billo,frau,giacone,lerda@to.infn.it; Francesco.Fucito,Francisco.Morales,Daniel.Ricci.Pacifici@roma2.infn.it} 
}
\abstract{
We derive the exact supergravity profile for the twisted scalar field
emitted by a system of fractional D3 branes at a $\mathbb Z_2$ orbifold singularity
supporting ${\cal N}=2$ quiver gauge theories with unitary groups and bifundamental matter.
At the perturbative level this twisted field is ``dual'' to the gauge coupling but it is corrected 
non-perturbatively by an infinite tower of fractional D-instantons.  The explicit microscopic description allows to derive the gravity profile from disk amplitudes computing 
the emission rate of the twisted scalar field in terms of chiral correlators 
in the dual gauge theory. We compute these quantum correlators using multi-instanton 
localization techniques and/or Seiberg-Witten analysis. Finally, we discuss a non-perturbative 
relation between the twisted scalar and the effective coupling of the gauge theory
for some simple choices of the brane set ups.}

\preprint{DFTT/07/2012\\
ROM2F/2012/03}
\begin{document}

\section{Introduction}

The embedding of gauge theories in a string framework by means of D-branes provides 
a remarkably simple and intuitive picture of the strong coupling dynamics of quantum field 
theories and of their non-perturbative phenomena.
For example, in a D-brane realization of a gauge theory, instantons are represented as
euclidean D-branes which are point-like in the four-dimensional space-time and aligned with the gauge
branes in the internal directions. The moduli parametrizing the general instanton solution 
are realized in terms of open strings with at least one end-point on the euclidean branes \cite{Witten:1995gx,Douglas:1995bn}. This is a very efficient and practical description
of instanton effects \cite{Green:2000ke,Billo:2002hm}.

On the other hand, the realization of gauge instantons as euclidean D-branes implies that 
they act also as sources of closed string fields, exactly as it happens for the D-branes supporting 
the gauge and matter degrees of freedom.
This is not surprising from a holographic point of view  since, in non-conformal settings, instantons generate a scale dependence in the gauge dynamics that must 
be reflected in the dual geometry
(for studies of gauge/gravity correspondences in non-conformal cases see for example the reviews 
\cite{Bigazzi:2003ui}\nocite{Bertolini:2003iv}--\cite{Nunez:2010sf} and references therein).
In searching for the gravitational solution dual to a non-conformal gauge theory, the starting point is
often given by the perturbative profile of the gravitational fields emitted by the brane system 
via disk diagrams. In presence of gauge instantons represented by euclidean branes,
there are extra source terms for the closed string fields and thus the supergravity solution 
corresponding to a given brane set-up gets modified with respect to its perturbative expression.
Incorporating explicitly all instanton corrections therefore provides the
microscopic derivation of the exact solution ``dual'' to the gauge theory,
yielding an extremely meaningful test of the gravitational description obtained using 
symmetry and/or duality considerations. It could also shed some light on aspects of the theory 
which have not yet been satisfactorily accommodated.

This line of thought is particularly fruitful in the context of $\mathcal{N}=2$ SYM
theories. In this case, the low-energy dynamics on the Coulomb branch 
is exactly described by a Seiberg-Witten (SW) curve 
\cite{Seiberg:1994rs,Seiberg:1994aj} that
summarizes in geometrical terms the infinite tower of multi-instanton
corrections to the low-energy effective action. These results can be
derived from first principles with the help of localization techniques that allow 
an explicit evaluation of integrals over the multi-instanton moduli space
\cite{Moore:1998et}\nocite{Nekrasov:2002qd,Flume:2002az,Bruzzo:2002xf,Nekrasov:2003rj}--\cite{Bruzzo:2003rw}. 
These techniques fit naturally in the brane context and generalize 
the SW results in many different directions  \cite{Billo:2006jm}\nocite{Ito:2008qe,Billo:2009di,Fucito:2009rs,Ito:2010vx}--\cite{Billo':2010bd}.
It is then natural to ask how the infinite tower of instanton effects is encoded in the holographic description of the gauge theory in terms of gravity.  This question has been addressed recently 
in Ref.s~\cite{Billo:2011uc}\nocite{Fucito:2011kb}--\cite{Billo:2012xj}, where
a four-dimensional SU(2) gauge theory was matched to a background generated 
by a stack of D7 branes on top of an O7 orientifold plane.
The dilaton-axion field governing the gauge coupling of the SU(2)
theory was derived from disk amplitudes computing the emission from O7, D7 and 
D-instanton sources, and the result was expressed in terms of chiral correlators in the
eight-dimensional gauge theory living on the D7/O7 system. 

In this paper we extend these methods to more general $\mathcal{N}=2$ SYM theories 
with SU($N$) gauge groups and fundamental matter. The set-up we consider involves 
fractional D3 branes of type IIB at non-isolated singularities, 
and in particular we take the $\mathbb{C}^2/\mathbb{Z}_2\times\mathbb{C}$ orbifold. 
By studying the emission of closed string fields from such branes, the corresponding perturbative 
supergravity solutions were constructed several years ago
\cite{Klebanov:1999rd}\nocite{Bertolini:2000dk,Polchinski:2000mx,Bertolini:2001qa,Petrini:2001fk}--\cite{Billo:2001vg}.
In these solutions a scalar field from the twisted sector, which we call
$t(z)$, varies logarithmically in the internal complex direction $z$
transverse to the orbifold and reproduces the
perturbative running of the gauge coupling with the scale. Such perturbative
solutions suffer from singularities at small values of $z$, {\it i.e.} in the IR
region of the gauge theory, and have to be modified by non-perturbative corrections.

The search for the corrected solutions has turned out to involve quite subtle
issues, such as the proposal that
the fractional branes originally placed at the origin inflate into an
enhan\c{c}on ring of size comparable to the dynamically generated scale
so as to avoid the IR singularity \cite{Johnson:1999qt} or the necessity 
of embedding the pure fractional brane solutions into a cascading theory of fractional 
plus regular branes \cite{Benini:2008ir}. Finally, the exact expression
of the twisted scalar $t(z)$ for a configuration of $N$ fractional
branes supporting a pure SU$(N)$ gauge theory was put forward in Ref.~\cite{Cremonesi:2009hq}.
This expression was found by exploiting the T-duality of the type IIB configuration to a type IIA
configuration of D4 branes suspended between NS5 branes, which can be uplifted
to M-theory \cite{Witten:1997sc}. In M-theory, the brane arrangement expands into 
the product of the four-dimensional spacetime and of a Riemann surface representing
the SW curve of the gauge theory, so that by uplifting the solution to M-theory one
incorporates the non-perturbative effects.

In this paper we show how the instanton corrections to the $t$ profile 
can be directly computed in the type IIB theory, without resorting to T-duality 
and the up-lift to M-theory. Our solution coincides, in the case
of pure gauge theories, with the one of Ref.~\cite{Cremonesi:2009hq} and generalizes the latter 
to the case when flavors are present or to quiver theories, thus providing a very deep test 
of the procedures and dualities involved in the supergravity derivation or, if we regard 
it the other way round, of the techniques we employ to explicitly compute such corrections. Furthermore, we will show that the field $t(z)$ can be identified, 
in the gauge/gravity dictionary, with the vacuum expectation value of a chiral operator ${\cal O}(z)$ in the four-dimensional
gauge theory which computes at the perturbative level the running  coupling of the gauge theory. 
At the exact level, instead, it contains the information on the
effective coupling matrix of the SU($N$) theory at the scale $z$ in a quite non-trivial fashion. We
shall derive this relation in terms of SW curves and illustrate it explicitly 
for the lowest-rank unitary groups. 

Let us now discuss a bit more in detail our approach and the structure of the paper.
In Section~\ref{secn:setup} we introduce the $\mathcal{N}=2$ quiver theory 
with gauge group $\mathrm{SU}(N_0)\times \mathrm{SU}(N_1)$ realized with fractional  
D3 branes in the $\mathbb{C}^2/\mathbb{Z}_2\times \mathbb C$ orbifold.  
We first review the derivation of the perturbative $t$ profile emitted by the D3 branes
and then show how it gets non-perturbatively modified by the fractional
D$(-1)$ branes. These modifications arise from the interactions between the instanton moduli
and the twisted scalar $t$, which we compute by evaluating disk amplitudes (Appendix \ref{secn:appa} 
contains a detailed derivation of the relevant string diagrams). 
The result can be written as
\begin{equation}
 \ii\pi  t(z) = \ii\pi t_0 -2\,
\Big\langle \tr_{N_0} \log \frac{z- \phi_0}{\mu} \Big\rangle
+2\,\Big\langle \tr_{N_1} \log \frac{z- \phi_1}{\mu} \Big\rangle
\label{tauz0}
\end{equation}
where $\phi_{0}$ and $\phi_1$ are the adjoint scalar fields in the vector multiplets of the two gauge 
groups and $\mu$ an arbitrary reference scale. Expanding both sides of (\ref{tauz0}) for large $z$ one finds a holographic
dictionary between derivatives of the gravity field $t(z)$ and chiral correlators in the four-dimensional gauge theory.
These correlators can be computed explicitly and quite efficiently 
order by order in the instanton expansion using the localization techniques
which are reviewed and summarized in Appendix~\ref{MultiInstanton}.
Neglecting the $\mathrm{SU}(N_1)$ dynamics, the quiver theory can be regarded as a 
$\mathrm{SU}(N_0)$ gauge theory with $2N_1$ fundamental flavors. In this case the correlators 
in (\ref{tauz0}) can be computed exactly to all orders in the instanton parameter
from the corresponding SW curve, and the $t$ profile can be written as
 \begin{equation}
 \ii\pi  t (z)= \log
 \frac{P(z)-\sqrt{ P^2(z) - g^2 Q(z)}}{P(z)+\sqrt{P^2(z) - g^2 Q(z)}}
\label{tsugra0}
\end{equation}
where $P$ and $Q$ are polynomials, respectively, of order $N_0$ and $N_1$ entering the 
SW curve and containing the 
vacuum expectation values and the mass parameters of the gauge theory, 
while $g^2$ is a specific function of the gauge coupling. 
This is the main result of Section~\ref{secn:chiral}.  

{From} the perturbative behavior it is natural to expect that 
the gravitational profile $t(z)$ encodes some information about the dynamics of the 
dual gauge theory at the scale $z$ and in particular about the matrix $\tau$ of its gauge 
couplings. The explicit relation between $t(z)$ and $\tau$ turns out 
to be non-trivial and quite interesting. In Section~\ref{secn:gaugegrav} we elaborate on this 
issue and consider the twisted field produced by the (instanton-corrected) 
fractional D3 branes placed at the origin of their quantum moduli space. In this configuration, which corresponds to the so-called ``enhan\c{c}on vacuum''
\cite{Benini:2008ir,Cremonesi:2009hq}, all gauge invariants $\vev{\tr\, \phi^k}$ 
constructed with the adjoint scalar field vanish and the gravitational profile of the twisted
field depends only on $z$. We then propose that such a profile encodes the 
effective coupling matrix $\tau$ of the gauge theory in the so-called ``special vacuum'' \cite{Argyres:1999ty} where all invariants 
$\vev{\tr\, \phi^k}$ but the highest vanish and the latter, namely 
$\mathbf{v} =  \frac{1}{N}\,\vev{\tr\, \phi^{N}}$, is related to the scale $z$ simply 
by $\mathbf{v} = z^N$~.
We show, by the comparison of SW curves, that the gauge theory at the special vacuum can be equivalently regarded as a massless conformal theory with a UV coupling $t$.
The explicit relation between $t$ and the gauge coupling is therefore the same which is found 
in conformal theories between the UV coupling (the one appearing in instanton computations \`a la Nekrasov) and the IR one. This relation  
depends on the rank of the group and involves modular functions. 
Our proposal is substantiated by explicit checks against the coupling matrices extracted from multi-instanton computations and  SW curves as shown in Appendix \ref{secn:appc} for the SU(2) and SU(3) theories. 

Finally, in Section \ref{secn:summary} we summarize our results, draw our conclusions and discuss some
possible developments of this work.

\section{The brane set-up and the $t$ profile}
\label{secn:setup}

\begin{figure}
\begin{center}
\begin{picture}(0,0)%
\includegraphics{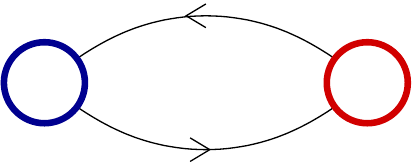}%
\end{picture}%
\setlength{\unitlength}{1657sp}%
\begingroup\makeatletter\ifx\SetFigFontNFSS\undefined%
\gdef\SetFigFontNFSS#1#2#3#4#5{%
  \reset@font\fontsize{#1}{#2pt}%
  \fontfamily{#3}\fontseries{#4}\fontshape{#5}%
  \selectfont}%
\fi\endgroup%
\begin{picture}(4706,1844)(1203,-3953)
\put(1441,-3121){\makebox(0,0)[lb]{\smash{{\SetFigFontNFSS{9}{10.8}{\rmdefault}{\mddefault}{\updefault}$N_0$}}}}
\put(5176,-3121){\makebox(0,0)[lb]{\smash{{\SetFigFontNFSS{9}{10.8}{\rmdefault}{\mddefault}{\updefault}$N_1$}}}}
\end{picture}%
\end{center}
\caption{The quiver diagram for the orbifold $\mathbb C^2/\mathbb Z_2$. The branes
of type 0 are represented by the blue circle on the left, while the branes of type 1 are represented
by the red circle on the right. The oriented lines connecting
the two types of branes represent the hypermultiplets in the bi fundamental representations.}
\label{fig:quiver}
\end{figure}
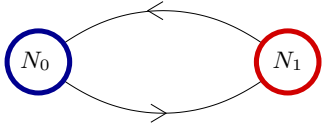

We study four-dimensional $\mathcal N=2$ SYM theories with unitary gauge groups realized with
fractional D3 branes at the non-isolated orbifold singularity 
$\mathbb C^2/\mathbb Z_2\times \mathbb C$. 
In this orbifold
there are two types of fractional D3 branes, which we call types 0 and 1,
corresponding to the two different irreducible representations
of the orbifold group $\mathbb Z_2$.
The most general brane configuration therefore consists of $N_0$ branes of type 0 and $N_1$
branes of type 1, and corresponds to a ${\mathcal N}=2$ quiver theory in four dimensions
with gauge group U($N_0$)$\times$U($N_1$), one hypermultiplet
in the bi fundamental representation $({\mathbf{\overline N_0}},{\mathbf{N_1}})$ and one hypermultiplet
in the $({\mathbf{N_0}},{\mathbf{\overline N_1}})$ representation. The corresponding quiver diagram is
represented in Fig.~\ref{fig:quiver}.
If one ignores the gauge degrees of freedom on the $N_1$ branes, {\it i.e.} turns off the
U($N_1$) gauge coupling and considers only the dynamics of the $N_0$ branes of type 0, one obtains
a ${\mathcal N}=2$ U($N_0$) SYM theory with $2N_1$ fundamental flavors and U($N_1$) as
global symmetry group. In this limit, with an abuse of language, we will sometimes refer to the
$N_0$ branes as the ``gauge'' or ``color'' branes and to the $N_1$
branes as the ``flavor'' branes.
Furthermore, we will decouple the U(1) factors and concentrate 
on the SU($N_0$)$\times$ SU($N_1$) part of the symmetry group.

We are interested in the Coulomb branch of the quiver gauge theory.
The degrees of freedom on this branch (vector multiplets) are described by the massless 
excitations of the open strings starting and ending on branes of the same type.  
They can be assembled in a chiral $\mathcal N=2$ superfield
$\Phi$ which is schematically given by
\begin{equation}
 \Phi= \phi + \theta \lambda +\frac{1}{2}\,\theta\,
\gamma^{\mu\nu}\tilde\theta\,F_{\mu\nu}+\cdots
\label{Phi}
\end{equation}
where the component fields $\phi$, $\lambda$, $F_{\mu\nu}$ are in the adjoint
representation of SU($N_0$)$\times$ SU($N_1$).  The superfield $\Phi$ can be written in the block diagonal form
\begin{equation}
\Phi=\begin{pmatrix} 
      \Phi_0& 0  \cr
     0&\Phi_1
     \end{pmatrix}
\end{equation}
where $\Phi_0$ and $\Phi_1$ are matrices in the adjoint representations of SU($N_0$) and SU($N_1$)
respectively. By giving a non-vanishing vacuum expectation value
to $\phi$ we can explore the Coulomb branch of the moduli space of the quiver gauge theory.
In the description where the dynamics of the D3 branes of type 1 is ignored, the field 
$\phi_1$ is frozen to its vacuum expectation value $\langle \phi_1 \rangle$ whose eigenvalues
parametrize the masses of the $2N_1$ fundamental flavors.

On the other hand, D-branes have a dual interpretation as gravitational sources, {\it i.e.} as
sources of closed string fields. Using for example the boundary state formalism as discussed in Ref.s~\cite{Bertolini:2000dk,Billo:2001vg,Bertolini:2001qa},
one can show that the fractional D3 branes in the $\mathbb Z_2$ orbifold are sources
of a non-trivial metric and a 4-form R-R potential from the untwisted sectors,
and of two scalars, $b$ and $c$, from the twisted NS-NS and R-R sectors respectively. While the
untwisted fields emitted by fractional D3 branes can propagate in all six transverse directions, 
the twisted scalars only propagate in the complex plane transverse to the D3 brane world-volume
which is not affected by the orbifold projection and which we parametrize with
a complex coordinate ${\mathbf x}$. Fractional D3 branes distributed on this plane therefore
generate a non-trivial dependence of the fields $b$ and $c$ on ${\mathbf x}$.
The twisted scalars are conveniently combined in a complex field
\begin{equation}
 t= c +\tau\,b
\label{t}
\end{equation}
where $\tau$ is the axio-dilaton of the type IIB string theory. For simplicity we assume that
the axion is trivial and that there are no branes other than the fractional D3 branes
so that the dilaton does not run. Thus, in this case we simply have $\tau=\ii/g_s$ where
$g_s$ is the string coupling constant. 
The equations of motion of type IIB supergravity require that $t$ be
a holomorphic function of ${\mathbf x}$.

The field $t$ is actually part of a chiral bulk superfield $T$
whose structure is schematically given by
\begin{equation}
 T=t+\cdots+\theta^4\, \frac{\partial^2 }{\partial {\bf x}^2} \, \bar t+\cdots
\label{T}
\end{equation}
with dots denoting supersymmetric descendants of $t$ and $\bar t$ being 
the complex conjugate of $t$. Fractional D3 branes and D-instantons are sources for the 
twisted closed string field $t$. This is the strict analogue of the set up with D7 
branes and O7 planes discussed in Ref.s~\cite{Billo:2011uc}\nocite{Fucito:2011kb}-\cite{Billo:2012xj},
where the axio-dilaton of type I supergravity was extracted from string amplitudes computing 
emission rates from D-brane and O-plane sources.

\subsection{The perturbative $t$ profile}
The profile of the twisted scalar $t$ emitted by the
fractional D3 branes described above 
can be derived by solving the classical field equations that
follow from the bulk action containing the kinetic terms and the source action
describing the emission from the fractional D3 branes. At the perturbative level this profile
was obtained long ago in Ref.s~\cite{Bertolini:2000dk,Polchinski:2000mx,Billo:2001vg,Bertolini:2001qa}. For completeness we now briefly review this derivation.
Let us first consider the NS-NS twisted scalar $b$ whose bulk action is
\begin{equation}
 S_{\mathrm{bulk}} = -\frac{(\pi^2\alpha')^2}{\kappa^2}\int d^6x\,
\Big(\frac{1}{2}\,\partial b\cdot\partial b
+\cdots\Big)~.
\label{sbulk1}
\end{equation}
Here $\kappa=8\pi^{{7}/{2}}\alpha'^2\,g_s$ is the gravitational coupling, and
the prefactor $(\pi^2\alpha')^2$ is due to our normalization conventions%
\footnote{We
adopt conventions such that $b$ and $c$ are dimensionless.}. Notice that
this bulk action is six-dimensional since the twisted fields can propagate only in the
six un-orbifolded directions. Then, let us consider the source action for $b$.
If all branes are at the origin, this is
\begin{equation}
  -\frac{T_3(N_0-N_1)}{2\kappa}\int d^6x \,
 b \,\delta^2({\bf x})
\label{source0}
\end{equation}
where $T_3=\sqrt{\pi}$ is the D3 brane tension. The appearance of the $\delta$-function indicates 
that the source term is localized on the world-volume of the D3 branes at
the origin.
By varying (\ref{source0}) and the bulk action (\ref{sbulk1}), we easily obtain the field equation
\begin{equation}
 \square\,
 b  = 4(N_0-N_1)g_s\,\delta^2({\mathbf x} )
\label{boxb}
\end{equation}
whose solution is
\begin{equation}
 b = b_0 \,+\,\frac{2(N_0-N_1)g_s}{\pi}\,\log \Big|\frac{{\mathbf x} }{{\mathbf x} _0}\Big|~.
\label{b0}
\end{equation}
Here $b_0=\frac{1}{2}$ is the classical value of the twisted scalar field in the $\mathbb Z_2$ orbifold
and $\mathbf x_0$ is an arbitrary length scale.
Adding the contribution of the scalar $c$ from the R-R twisted sector, which is proportional to
the argument of $\mathbf x$ \cite{Bertolini:2000dk}, we obtain the following holomorphic solution
\begin{equation}
 \ii\pi t = \ii\pi t_0-2(N_0-N_1) \log\frac{ {\mathbf x}   }{ {\mathbf x} _0}
\label{proft}
\end{equation}
with $t_0= \ii\, b_0/g_s$. It is convenient to introduce the quantities
\begin{equation}
 z =\frac{  {\bf x}   }{2\pi\alpha'}~~~~\mbox{and}~~~~\mu=\frac{  {\bf x} _0}{2\pi\alpha'}
\label{ymu}
\end{equation}
with mass dimension 1, and rewrite the solution (\ref{proft}) as follows
\begin{equation}
 \ii\pi t(z) = \ii\pi t_0-2(N_0-N_1) \log\frac{ z }{ \mu }~.
\label{proft1}
\end{equation}
In the conformal case, {\it i.e.} when $N_0=N_1$, we simply have
\begin{equation}
 t(z) = t_0 ~.
\label{proftconf}
\end{equation}

The profile (\ref{proft1}) has the correct logarithmic behavior to be identified
with the 1-loop running coupling constant at a scale $z$ of a SU($N_0$) gauge theory with $2N_1$
fundamental flavors Ref.s~\cite{Bertolini:2000dk,Polchinski:2000mx,Billo:2001vg,Bertolini:2001qa}.
We are therefore led to the following gauge/gravity relation
\begin{equation}
 t ~ \longleftrightarrow ~\tau_{\mathrm{gauge}} \equiv \frac{\theta_{\mathrm{YM}}}{\pi}+
\ii\,\frac{8\pi}{g^2_{\mathrm{YM}}}
\label{taut}
\end{equation}
where $g_{\mathrm{YM}}$ and ${\theta_\mathrm{YM}}$ are the coupling constant and the $\theta$-angle of
the SU($N_0$) SYM theory.
We will see, however, that when non-perturbative
effects are taken into account the relation between $t$ and $\tau_{\mathrm{gauge}}$ is more complicated than that in (\ref{taut}).

Let us now consider a more general configuration in which the D3 branes are not all at the
origin. This corresponds to giving non-vanishing vacuum expectation values to the adjoint
scalars $\phi_0$ and $\phi_1$, namely
\begin{equation}
 \begin{aligned}
\langle \phi_0 \rangle = \mathrm{diag}(a_1,\ldots,a_{N_0})~~~\mbox{and}~~~
   \langle \phi_1 \rangle  = \mathrm{diag}(b_1,\ldots,b_{N_1})
 \end{aligned}
 \label{am}
\end{equation}
with $\sum_u a_u=0$ and $\sum_{u'}b_{u'}=0$.
Repeating the same steps as before, one can show that the $t$ profile corresponding to such configuration is
 \begin{equation}
 \begin{aligned}
  \ii\pi t(z)
           &=\ii\pi t_0-2\,\tr_{ N_0 }\log \frac{z-\langle \phi_0\rangle}{\mu}
+2\,\tr_{N_1}\log \frac{z-\langle \phi_1\rangle}{\mu}
\end{aligned}
\label{proftam}
\end{equation}
where $\tr_{N_0}$ and $\tr_{N_1}$ denote the traces over the
SU($N_0$) and SU($N_1$) indices respectively.
Notice that when $\langle \phi_0\rangle$ and/or $\langle \phi_1\rangle$ are non-zero, 
the twisted scalar $t$ is not trivial even for $N_0=N_1$ since the conformal symmetry is broken.

In the following we concentrate on the case $N_0=N_1$ from which all others can be retrieved
by taking suitable decoupling limits in which subsets of branes are sent to infinity.
It is not difficult to realize that in this case $t$ satisfies
the following differential equation
\begin{equation}
\square\,t= 8 J_{\mathrm{cl}}\,\delta^2(z)
\label{tcl00}
\end{equation}
with
\begin{equation}
 J_{\mathrm{cl}}= \sum_{\ell=1}^\infty
\frac{\ii}{\ell\,!}\,\Big(\tr_{N_0} \langle \phi_0\rangle^{\ell} - \tr_{N_1} \langle \phi_1 \rangle^{\ell}\Big)
 \,\frac{\partial^{\ell}}{\partial z^{\ell}}
~=~ \ii\Big(\tr_{N_0} \,\ee^{\ii\,\bar p\,\langle\phi_0\rangle}
- \tr_{N_1} \,\ee^{\ii\,\bar p\,\langle \phi_1\rangle}\Big)
 \label{tcl0}
 \end{equation}
where in the second step we exploited the fact that $N_0=N_1$ and introduced the momentum operator conjugate to $z$, namely $\bar p = -\ii\partial/\partial z$.

The current $J_{\mathrm{cl}}$ has a nice interpretation in terms of disk diagrams describing
the couplings among the closed string twisted fields and the massless open string
excitations of the fractional D3 branes. To see this, let us consider the NS-NS scalar $b$
whose vertex operator we denote by $V_b$, and study its interactions with the scalar
$\phi_0$ of the type 0 branes whose vertex we denote by $V_{\phi_0}$.
When the scalars are frozen to their vacuum expectation values,
these interactions are given by
\begin{equation}
\sum_{\ell=0}^\infty \frac{1}{\ell\,!}\,\big\langle
\underbrace{V_{\phi_0}\cdots V_{\phi_0}}_{\ell}\,V_b\,\big\rangle_{\mathrm{D3}_0}
~=~ \frac{\pi}{g_s}\sum_{\ell=0}^\infty \frac{1}{\ell\,!}\,\tr_{N_0}\langle{\phi_0}\rangle^\ell\,(\ii\bar p)^\ell\,b
~=~\frac{\pi}{g_s}\,\tr_{N_0} \,\ee^{\ii\,\bar p\,\langle\phi_0\rangle}\,b
\label{Aell}
\end{equation}
where the factorial has been introduced as a symmetry factor and the right hand side
follows by computing the correlation functions of the vertex operators using
standard CFT techniques as discussed for example in Ref.~\cite{Billo:2011uc}.
A completely similar calculation can be performed with the scalar $\phi_1$ of the type 1 branes
leading to
\begin{equation}
 -\frac{\pi}{g_s}\,\tr_{N_1} \,\ee^{\ii\,\bar p\,\langle \phi_1\rangle}b\,~.
\end{equation}
where the extra sign comes from the fact that branes of type 1 have opposite $b$-charge with
respect of those of type 0.
Adding an analogous term describing the interactions of the R-R twisted scalar $c$, we can write the total contribution to the effective action (which is minus the scattering amplitude) as
\begin{equation}
 -\ii\pi\Big(
 \tr_{N_0} \,\ee^{\ii\,\bar p\,\langle\phi_0\rangle}-
\tr_{N_1} \,\ee^{\ii\,\bar p\,\langle \phi_1\rangle}
\Big)\bar t~.
\label{intert}
\end{equation}
Supersymmetry requires that the interaction term (\ref{intert}) must be accompanied by other
structures (that could also be computed from string diagrams with extra fermionic insertions)
in such a way that the effective action follows from a holomorphic
prepotential. As discussed in Ref.s~\cite{Billo:2011uc,Billo:2012xj}, such a prepotential
can be obtained simply by promoting the bulk and boundary scalars appearing in (\ref{intert})
to the corresponding chiral superfields. Denoting by $\delta T$ the fluctuation part of
$T$, we then find that the classical prepotential
contains the following term
\begin{equation}
\delta F_{\mathrm{cl}} = \ii\pi\Big(
\tr_{N_0} \,\ee^{\ii\,\bar p\,\Phi_0}-\tr_{N_1} \,\ee^{\ii\,\bar p\, \Phi_1}
\Big)\frac{\delta T}{\bar p^2}+\cdots
\label{F}
\end{equation}
where the dots represent interactions of higher orders in $\delta T$.
Note that this expression is well-defined for $\bar p\to 0$ since for $N_0=N_1$
the traces inside brackets start to contribute at order $\bar p^2$.
The effective action follows upon integrating the prepotential over $d^4\theta$;
when all four $\theta$'s are taken from $\delta T$ and the superfields $\Phi_0$ and
$\Phi_1$ are frozen to their vacuum expectation values, we recover the interaction
(\ref{intert}). This is a source term for $t$ corresponding to the classical
current (\ref{tcl0}) which is related to the prepotential as follows
\begin{equation}
 J_{\mathrm{cl}}= \frac{\bar p^2}{\pi}\,
\frac{\delta F_{\mathrm{cl}}}{ \delta T}\Bigg|_{\Phi\to \langle \Phi \rangle}~.
\label{jclprep}
\end{equation}

\subsection{The non-perturbative $t$ profile}

Let us now investigate how the classical profile (\ref{proftam})
changes when non-perturbative effects due to gauge instantons are taken into account.
In our brane set-up, instantons  are introduced by adding $k_0$ fractional D(--1) branes of
type 0 and $k_1$ fractional D(--1) branes of
type 1. D-instantons of type 0 correspond to gauge instantons for the gauge group SU($N_0$) while
they are ``exotic" from the perspective of the SU($N_1$) gauge group. The situation gets reversed for the D-instantons of type 1. In the case where the gauge dynamics of the fractional D3 branes
of type 1 is ignored, instantons of type 1 are consistently discarded.

The physical excitations of the open strings with at least one end-point on
D(--1) branes account for the instanton moduli which we
collectively denote as $\cM_{k}$ with $k=(k_0,k_1)$.
They comprise the neutral sector, corresponding to D(--1)/D(--1) open strings
that do not transform under the gauge groups,
and the sectors arising from D(--1)/D3$_0$ and
D(--1)/D3$_1$ open strings that transform in the fundamental (or anti-fundamental)
representations of the gauge groups.
The complete list of instanton moduli and their transformation properties are given in 
Appendix~\ref{MultiInstanton}. 
Here we just recall that among the neutral moduli we have the
bosonic and fermionic Goldstone modes of the supertranslations of the
D3 world-volume which are broken by the D-instantons and which are then
identified with the superspace coordinates $x$ and
$\theta$, and a pair of complex scalars $\chi_{0}$ and $\chi_{1}$ (plus their conjugates $\bar\chi_{0}$ and $\bar\chi_{1}$) transforming in the adjoint representations of the instanton symmetry groups 
U($k_0$) and U($k_1$) respectively. The eigenvalues of these matrices describe the position
of the D-instantons in the un-orbifolded directions transverse to the fractional D3 branes
and are the D(--1) analogue of $\langle\phi_{0}\rangle$ and $\langle\phi_{1}\rangle$.

In order to find the non-perturbative $t$ profile we first compute the instanton induced prepotential
$F_{\mathrm{n.p.}}$ from which the non-perturbative source current $J_{\mathrm{n.p.}}$ can be
derived following a procedure similar to the one outlined for the classical current $J_{\mathrm{cl}}$.
The non-perturbative prepotential is defined as
\begin{equation}
 F_{\mathrm{n.p.}}= \sum_k \int \!\!d\widehat{\mathcal M}_{k}
~\ee^{-S_{\mathrm{inst}}({\mathcal M}_{k}, \Phi, T)}
\label{Fnp}
\end{equation}
where the integral is performed over the centered moduli $\widehat{\mathcal M}_{k}$, including
all moduli except the superspace coordinates $x$ and $\theta$. 
$S_{\mathrm{inst}}({\mathcal M}_{k}, \Phi, T)$ is the moduli action, describing
the interactions of the instanton moduli with the boundary and bulk superfields.

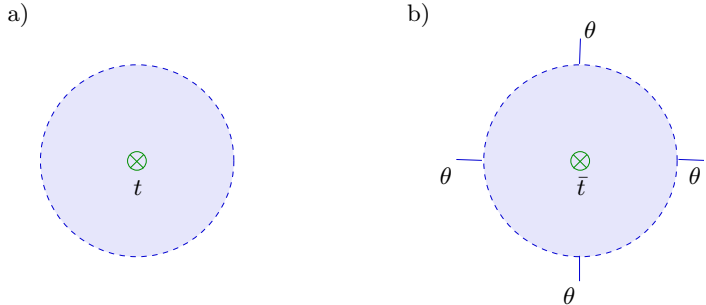
\begin{figure}
\begin{center}
\begin{picture}(0,0)%
\includegraphics{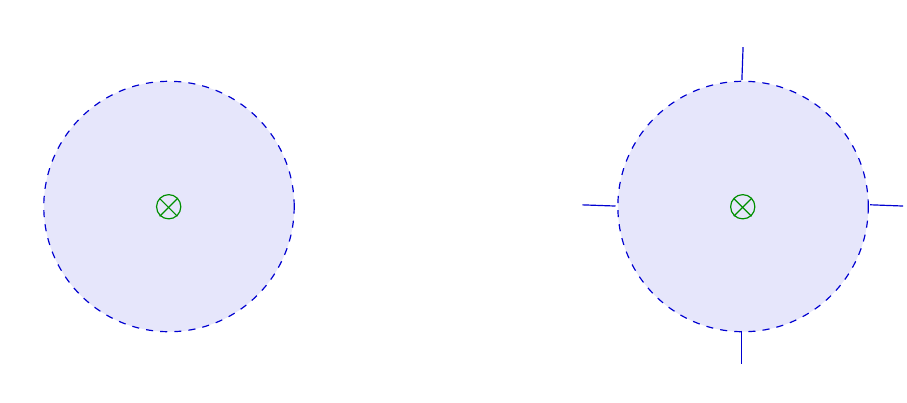}%
\end{picture}%
\setlength{\unitlength}{1492sp}%
\begingroup\makeatletter\ifx\SetFigFontNFSS\undefined%
\gdef\SetFigFontNFSS#1#2#3#4#5{%
  \reset@font\fontsize{#1}{#2pt}%
  \fontfamily{#3}\fontseries{#4}\fontshape{#5}%
  \selectfont}%
\fi\endgroup%
\begin{picture}(11489,5106)(121,-4369)
\put(136,434){\makebox(0,0)[lb]{\smash{{\SetFigFontNFSS{9}{10.8}{\familydefault}{\mddefault}{\updefault}a)}}}}
\put(6706,434){\makebox(0,0)[lb]{\smash{{\SetFigFontNFSS{9}{10.8}{\familydefault}{\mddefault}{\updefault}b)}}}}
\put(2206,-2446){\makebox(0,0)[lb]{\smash{{\SetFigFontNFSS{9}{10.8}{\familydefault}{\mddefault}{\updefault}$t$}}}}
\put(9616,164){\makebox(0,0)[lb]{\smash{{\SetFigFontNFSS{9}{10.8}{\familydefault}{\mddefault}{\updefault}$\theta$}}}}
\put(7246,-2266){\makebox(0,0)[lb]{\smash{{\SetFigFontNFSS{9}{10.8}{\familydefault}{\mddefault}{\updefault}$\theta$}}}}
\put(9496,-2446){\makebox(0,0)[lb]{\smash{{\SetFigFontNFSS{9}{10.8}{\familydefault}{\mddefault}{\updefault}$\bar t$}}}}
\put(9271,-4246){\makebox(0,0)[lb]{\smash{{\SetFigFontNFSS{9}{10.8}{\familydefault}{\mddefault}{\updefault}$\theta$}}}}
\put(11341,-2266){\makebox(0,0)[lb]{\smash{{\SetFigFontNFSS{9}{10.8}{\familydefault}{\mddefault}{\updefault}$\theta$}}}}
\end{picture}%
\end{center}
\caption{a) The disk diagram describing the interaction of the D-instantons with $t$.
b) The diagram, linked by supersymmetry to the previous one, which
comprises the insertion of four vertices for $\theta$ and one vertex for $\bar t$.}
\label{fig:ttheta}
\end{figure}

In our stringy set-up the moduli action is derived by computing string amplitudes
on disks with at least a portion of their boundary lying on the D(--1) branes as discussed in detail
in Ref.s~\cite{Green:2000ke,Billo:2002hm}. Here we only discuss the dependence of
$S_{\mathrm{inst}}$ on the twisted bulk superfield $T$, since the other part is quite
standard and can be found in the literature.
The simplest term involving $T$ corresponds to the coupling of the twisted scalars
to the D-instantons which can be obtained from the
Born-Infeld and Wess-Zumino actions of $k$ fractional D(--1) branes, namely
\begin{equation}
   - \ii \pi k_0 t -\ii\pi  k_1 (\tau-t)~.
\label{S01a}
\end{equation}
Equivalently, (\ref{S01a}) can be computed as the sum of the classical instanton action
$S_{\mathrm{cl}}$ and the overlap between the boundary state of $k$ D-instantons
and the vertex operators for the twisted scalars $b$ and $c$ as discussed in Ref.s~\cite{Bertolini:2000dk,Billo:2001vg,Bertolini:2001qa}. 
The string amplitude corresponding to (\ref{S01a}) is represented
by the disk diagram of Fig.~\ref{fig:ttheta}a,
with the two terms coming from insertions on disks of type 0 and 1
respectively%
\footnote{Recall that, classically, the complex gauge 
couplings are $t$ and $\tau-t$ for the branes of type 0 and 1, respectively. Taking into account the background value $b_0$ of the NS-NS twisted scalar, we can rewrite (\ref{S01a}) as
$ - \ii\pi  \big[b_0k_0+(1-b_0)k_1\big]\,\tau- \ii\pi (k_0-k_1)\,\delta t~,$
where $\delta t$ is the fluctuating part of $t$.}. 
When $t$ reduces to $t_0=b_0 \tau= \ft12\tau$ at the orbifold point,
(\ref{S01a}) reduces to the classical instanton action $S_{\mathrm{cl}}=
-  \ii \pi  \,(k_0+k_1)\,t_0 $.
Thus, the $k$-instanton contributions to the prepotential (\ref{Fnp}) are weighted by
\begin{equation}
 \ee^{-S_{\mathrm{cl}} } = q_0^{k_0} \, q_1^{k_1}
\end{equation}
with
\begin{equation}
q_0=\ee^{\ii\pi t_0}=\ee^{-\frac{\pi b_0}{g_s}}~~~~\mbox{and}~~~~
q_1=\ee^{\ii\pi (\tau-t_0)}=\ee^{-\frac{\pi (1-b_0)}{g_s}}~.
\label{q0}
\end{equation}
Actually, by exploiting the broken supersymmetries we can promote $t$ to the full-fledged superfield
$T$, so that (\ref{S01a}) becomes
\begin{equation}
-\ii\pi k_1\,\tau-  \ii\pi  (k_0-k_1) \,T= S_{\mathrm{cl}}
-  \ii\pi (k_0-k_1) \,\delta T~.
\label{S01b}
\end{equation}
This action includes, among others, 
the structure $\theta^4\,\bar p^2 \, {\bar t}$
that is represented by the diagram in Fig.~\ref{fig:ttheta}b (see Appendix \ref{secn:appa}
for details on the string derivation of this interaction).

In presence of a non-constant $t$ we expect further interaction terms in
the moduli action involving the derivatives of $t$. In particular, we expect to find
the D-instanton analogues of the interactions (\ref{Aell}), namely
the coupling of the twisted scalars $b$ and $c$ with an arbitrary number of $\chi$ moduli 
whose vertex operator we denote by $V_\chi$. 
Indeed, as shown in Appendix \ref{secn:appa}, for insertions on a disk of type 0
we have
\begin{equation}
\sum_{\ell=0}^\infty
\frac{1}{\ell\,!}\,\big\langle
\underbrace{V_{\chi_0} \cdots V_{\chi_0} }_{\ell}\,V_b\big\rangle_{\mathrm{D}(-1)_0}
~=~ -\frac{\pi}{g_s}\sum_{\ell=0}^\infty\frac{1}{\ell\,!}\,\tr_{k_0}\chi_0^\ell\,(\ii\bar p)^\ell\,b
~=~-\frac{\pi}{g_s}\,\tr_{k_0}\ee^{\ii\bar p \,\chi_0} \,b
\label{chiell}
\end{equation}
where $\tr_{k_0}$ stands for the trace over the $k_0$ D-instantons of type 0.
In a similar way we can compute the interaction of the R-R scalar $c$ with the $\chi$ moduli. Adding this to the previous amplitude, we obtain
the following contribution to the instanton moduli action
\begin{equation}
 -\ii\pi\,\tr_{k_0}\ee^{\ii\bar p \,\chi_0}\,t~.
\label{tchi}
\end{equation}
By exploiting the broken supersymmetries or, equivalently,
by inserting vertices for the superspace coordinates $\theta$, we can
promote $t$ to the complete superfield $T$. In this way we
obtain, among others, the  terms
$\tr_{k_0}\ee^{\ii\bar p \,\chi_0} ~\theta^4\,\bar p^2\,{\bar t}$, 
which are responsible for the non-perturbative corrections in the $t$ profile.
Similar results (with the opposite sign) are found for insertions on instanton disks of type 1.
Collecting all contributions, we conclude that the action for $k$
fractional D-instantons in a non-trivial
twisted background is
\begin{equation}
S_{\mathrm{inst}}(\cM_{k}, \Phi, T) = S_{\mathrm{cl}}+ S'_{\mathrm{inst}}(\cM_{k}, \Phi)
-\ii\pi~\Big(\tr_{k_0}\ee^{\ii\bar p \,\chi_0}
- \tr_{k_1}\ee^{\ii\bar p \,\chi_1} \Big) \,\delta T+\ldots
\label{sinst1}
\end{equation}
where $S'_{\mathrm{inst}}$ is the part of the moduli action accounting for the interactions
of the instanton moduli among themselves and with the fields in the vector multiplet.

Using the moduli action (\ref{sinst1}) in (\ref{Fnp}) we can find how the non-perturbative prepotential
depends on the twisted scalar and extract from it the non-perturbative source current for $t$. 
To linear order in $\delta T$ one gets
\begin{equation}
 \delta F_{\mathrm{n.p.}} = \ii\pi\,\delta T \, \sum_k q_0^{k_0} q_1^{k_1}
\int \!\!d\widehat{\mathcal M}_{k}
~\ee^{-S'_{\mathrm{inst}}({\mathcal M}_{k}, \Phi)}\,\Big(\tr_{k_0}\ee^{\ii\bar p \,\chi_0}
- \tr_{k_1}\ee^{\ii\bar p \,\chi_1} \Big)~.
\label{Fnp2}
\end{equation}
In order to perform the integration over the moduli space and
obtain explicit results, we exploit the localization formula and adopt
Nekrasov's approach to the multi-instanton calculus Ref.s~\cite{Nekrasov:2002qd,Nekrasov:2003rj}
(see also Ref.s~\cite{Moore:1998et,Bruzzo:2002xf,Flume:2004rp}).
For details, we refer the reader to Ref.s~\cite{Billo:2011uc,Fucito:2011kb,Billo:2012xj} 
where similar manipulations have been performed for
D(--1)/D3/D7 systems in orientifold models and to Appendix \ref{MultiInstanton}. 
Here we just recall the
relation between the integrals in the right hand side of (\ref{Fnp2}) and the chiral correlators
in the gauge theory. Such a relation can be derived from the localization procedure in which
one first defines the deformed instanton partition function 
\begin{equation}
 Z_{\mathrm{inst}}= \sum_k q_0^{k_0} q_1^{k_1}\!
  \int \!\!d {\mathcal M}_{k}
~\ee^{-S'_{\mathrm{inst}}({\mathcal M}_{k}, \Phi; \epsilon_1,\epsilon_2)}  
\label{Zinst}
\end{equation}
and then the prepotential
\begin{equation}
 F_{\mathrm{n.p.}}= -\lim_{\epsilon_1,\epsilon_2 \to 0}
\epsilon_1 \epsilon_2  \log  Z_{\mathrm{inst}}~.
\label{prepotinst}
\end{equation}
In these expressions $\epsilon_1$ and $\epsilon_2$ are deformation parameters which in our string set-up
can be introduced by putting the brane system in a graviphoton background
Ref.s~\cite{Billo:2006jm,Ito:2010vx}. Then one can show that
\begin{equation}
\begin{aligned}
\frac{1}{(\ell+2)!}\,\Big\langle
\tr_{N_0}  \phi_0^{\ell+2} \Big\rangle_{\!\mathrm{inst}} \!\!&=-\frac{1}{\ell\,!}
\lim_{\epsilon_1,\epsilon_2\to 0}
\frac{\epsilon_1 \epsilon_2}{Z_{\mathrm{inst}}} 
\sum_k q_0^{k_0} q_1^{k_1}\!
  \int \!\!d {\mathcal M}_{k}
~\ee^{-S'_{\mathrm{inst}}({\mathcal M}_{k}, \Phi;\epsilon_1,\epsilon_2)}\,
\tr_{k_0}\chi_0^{\ell}\\
\end{aligned}
\label{trphi0}
\end{equation}
and similarly for the correlators on the type 1 branes with $\tr_{N_0}  \phi_0^{\ell+2}$ and
$\tr_{k_0}\chi_0^{\ell}$ replaced by $\tr_{N_1}  \phi_1^{\ell+2}$ and
$\tr_{k_1}\chi_1^{\ell}$ respectively.
Notice that the integrals in Eq.s~(\ref{Zinst}) and (\ref{trphi0}) are over all
moduli including $x$ and $\theta$, and that in the limit $\epsilon_i \to 0$ the factor $\epsilon_1 \epsilon_2$ in (\ref{trphi0}) compensates for the volume $V\sim \frac{1}{\epsilon_1\epsilon_2}$ of the regularized four dimensional superspace.
Plugging the relations (\ref{trphi0}) into (\ref{Fnp2}) one gets
\begin{equation}
 \delta F_{\mathrm{n.p.}}  = \ii\pi\,\Big\langle
\tr_{N_0} \,\ee^{\ii\,\bar p\,\Phi_0}-\tr_{N_1} \,\ee^{\ii\,\bar p\,\Phi_1}
\Big\rangle_{\!\mathrm{inst}} ~ \frac{\delta T}{\bar p^2}
\label{Fnp3}
\end{equation}
which is nothing but the instanton completion of (\ref{F}). 

Summarizing, we have found
that the instanton corrections to the prepotential yield a source
current for $t$ which has the same expression as the classical current (\ref{tcl0}) but with
the classical gauge invariants $\tr_{N_a}\big\langle \phi_a\big\rangle^\ell$,
replaced by the full quantum gauge invariants $\big\langle\tr_{N_a} \phi_a^\ell\big\rangle$.
The latter are computed by taking into account the complete dynamics on the gauge branes and
contain explicit non-perturbative corrections induced by the D-instantons.
We stress that the quantum correlators $\big\langle\tr_{N_0} \phi_0^\ell\big\rangle$ 
(and the same for the correlators of type 1) receive contributions 
from instantons of both types since the instanton action depends on both $\chi_0$ and $\chi_1$.

Finally, adding the classical and the instanton contributions we obtain the full source current 
for $t$:
\begin{equation}
 J= \frac{\bar p^2}{\pi}\,\frac{\delta F}{ \delta T}   
=\ii\,\Big\langle \tr_{N_0}  \,\ee^{\ii\,\bar p\,\phi_0}
 -\tr_{N_1} \,\ee^{\ii\,\bar p\, \phi_1}\Big\rangle
\label{Jex}
\end{equation}
where $\delta F=\delta F_{\mathrm{cl}}+\delta F_{\mathrm{n.p.}}$.
The field equation for $t$ in configuration space is therefore
\begin{equation}
 \square\, t= 8 J\,\delta^2(z)~=~8
\sum_{\ell=0}^\infty
\frac{\ii}{\ell\,!}\,\Big\langle \tr_{N_0}  \phi_0^{\ell} - \tr_{N_1}  \phi_1^{\ell} \Big\rangle   \,\frac{\partial^{\ell}}{\partial z^{\ell}}\,\delta^2(z)
\label{fecom}
\end{equation}
which is solved by
\begin{equation}
 \ii\pi  t(z) = \ii\pi  t_0 -2\,
\Big\langle \tr_{N_0} \log \frac{z- \phi_0}{\mu} \Big\rangle
+2\,\Big\langle \tr_{N_1} \log \frac{z- \phi_1}{\mu} \Big\rangle
~.
\label{tauz}
\end{equation}
This explicit solution shows that the non-trivial information about the $t$ profile
is contained in the ring of chiral correlators of the quiver gauge theory
which accounts for the full tower of D-instanton corrections to the gravity solution.
The appearance of the chiral correlators of a gauge theory in a gravitational profile
was observed in Ref.~\cite{Billo:2010mg} in an orientifold model with D7 branes and
explicitly checked at the first few instanton numbers in Ref.s~\cite{Billo:2011uc,Billo:2012xj}
and proved in full generality in Ref.~\cite{Fucito:2011kb}. Our current results provide another
example of this non-trivial gauge/gravity relation in an orbifold set-up.

We conclude by observing that the pure SU($N_0$) theory can be obtained from the conformal
one by decoupling all flavors. Alternatively, one could start from the beginning
simply with a stack of $N_0$ D3 branes of type 0 without any branes of type 1.
In either case one finds
\begin{equation}
 \ii\pi   t(z) = -2\,
\Big\langle \tr_{N_0} \log \frac{z- \phi_0}{\Lambda} \Big\rangle
\label{taupure}
\end{equation}
where $\Lambda$ is the dynamically generated scale of the theory: $\Lambda^{2N_0}=\ee^{\ii\pi t_0}\,\mu^{2N_0}~.$

\section{The gauge theory description of the $t$ profile}
\label{secn:chiral}

In this section we compute the gauge correlator (\ref{tauz}) determining the supergravity $t$ profile.
This calculation can be performed in an explicit way order by order in the instanton expansion 
using Nekrasov's approach to the multi-instanton calculus and the localization techniques 
which are reviewed in Appendix~\ref{MultiInstanton}. However, in the limit $q_1\to0$ where
the dynamics of the branes of type 1 is ignored, instead of an order-by-order instanton calculation 
we can use the SW curve for the effective SU($N_0$) gauge theory with $2N_1$ fundamental flavors 
and extract from it the quantum correlators in a closed and exact form. This is
what we do first in the following subsection.
Next we consider the quiver theory with gauge group $\mathrm{SU}(N_0)\times\mathrm{SU}(N_1)$. 
For $N_0=N_1$, the SW curve for the quiver theory 
was derived in Ref.~\cite{Witten:1997sc}, where the model was engineered in terms of two 
stacks of $N_0$ and $N_1$ D4-branes stretching between two NS5 branes living on a circle 
(see also Ref.~\cite{Petrini:2001fk}).
Yet a microscopic derivation of this curve and a detailed quantitative study of the chiral 
correlators in the quiver gauge theory are missing. Therefore, in this case we apply Nekrasov's multi-instanton calculus to derive the correlator determining the 
$t$ profile order by order in $q_0$ and $q_1$. 

\subsection{The exact $t$ profile from the Seiberg-Witten curve for SU($N$)}
\label{secn:chiral1}

We start by considering the case where the dynamics of branes of type 1 is ignored.  In this limit we can replace everywhere the field $\phi_1$ by its vacuum expectation value
\begin{equation}
\langle \phi_1 \rangle  ={\rm diag} (m_1,\cdots, m_{N_1} )~.
\end{equation}
As compared to (\ref{am}), here we have denoted by $m$'s the eigenvalues of $\langle\phi_1\rangle$,
since in this case they correspond to masses for the fundamental hypermultiplets.
On the other hand the field $\phi_0$, which in this section we simply refer to as $\phi$,
is a dynamical field spanning the Coulomb branch of the SU($N_0$) gauge theory. 
For simplicity we take $N_0=N_1\equiv N$, but the results for $N_0>N_1$ can be recovered by
decoupling some of the hypermultiplets by sending some of the $m$'s to infinity. 

The chiral correlators $\big\langle \tr\,\phi^\ell\big\rangle$ can be computed from
(\ref{trphi0}) using Nekrasov's approach to the multi-instanton calculus. 
Equivalently (and more efficiently) they can be obtained from the SW curve associated to 
a SU($N$) SYM theory with $N_f=2N$ fundamental flavors with masses%
\footnote{Recall that in our set-up the flavors are realized by means of fractional D3 branes of type 1 and that the hypermultiplets correspond to open strings
stretching between the gauge and the flavor branes. For each one of these we have then two hypermultiplets with same mass corresponding to the two orientations of the open strings. If one wants to have $N_f$ generically distinct masses one has to realize the flavors by means of $N_f$
fractional D7 branes of type 0 as done for example in Ref.~\cite{Bertolini:2001qa}.}
\begin{equation}
 \{m_f\}=\{m_1,\cdots,m_{N},m_1,\cdots,m_{N}\}~.
\label{masses}
\end{equation}
Such a curve can be written as \cite{Argyres:1999ty}
\begin{equation}
 y^2= P(z)^2 - g^2\, Q(z)
\label{curve}
\end{equation}
where
\begin{equation}
 \begin{aligned}
  P(z) = \prod_{u=1}^{N} (z-e_u) \,= \sum_{\ell=0}^N u_\ell\, z^{N-\ell}~,~~~~
 Q(z) = \prod_{u=1}^{N}  (z-m_u)^2\,=\Big( \sum_{\ell=0}^{N}  T_{\ell}\, z^{N-\ell}\Big)^2~,
 \end{aligned}
\label{PQ}
\end{equation}
and
\begin{equation}
 g^2 =  \frac{4q_0}{(1+q_0)^2} 
\label{g2}
\end{equation}
with $q_0=\ee^{\ii\pi t_0}$. The parameters $e_u$ and $m_u$ are subject to the constraints
\begin{equation}
\sum_{u=1}^{N} e_u=0\qquad\mbox{and}\qquad\sum_{u=1}^N m_u=0~.
\label{sumem}
\end{equation}
The first condition is related to the fact that we have a SU($N$) gauge symmetry rather than
a U($N$) one, while the restriction on the $m$'s is just for the sake of simplicity (see also (\ref{am})). 
Therefore we can take the first $N-1$ $e$'s and $m$'s as independent variables, which we will label
by an index $i=1,\cdots,N-1$.

\begin{figure}
    \begin{center}
    \def\svgwidth{7.5cm}
    \begingroup%
  \makeatletter%
  \providecommand\color[2][]{%
    \errmessage{(Inkscape) Color is used for the text in Inkscape, but the package 'color.sty' is not loaded}%
    \renewcommand\color[2][]{}%
  }%
  \providecommand\transparent[1]{%
    \errmessage{(Inkscape) Transparency is used (non-zero) for the text in Inkscape, but the package 'transparent.sty' is not loaded}%
    \renewcommand\transparent[1]{}%
  }%
  \providecommand\rotatebox[2]{#2}%
  \ifx\svgwidth\undefined%
    \setlength{\unitlength}{360.01045375bp}%
    \ifx\svgscale\undefined%
      \relax%
    \else%
      \setlength{\unitlength}{\unitlength * \real{\svgscale}}%
    \fi%
  \else%
    \setlength{\unitlength}{\svgwidth}%
  \fi%
  \global\let\svgwidth\undefined%
  \global\let\svgscale\undefined%
  \makeatother%
  \begin{picture}(1,0.71111536)%
    \put(0,0){\includegraphics[width=\unitlength]{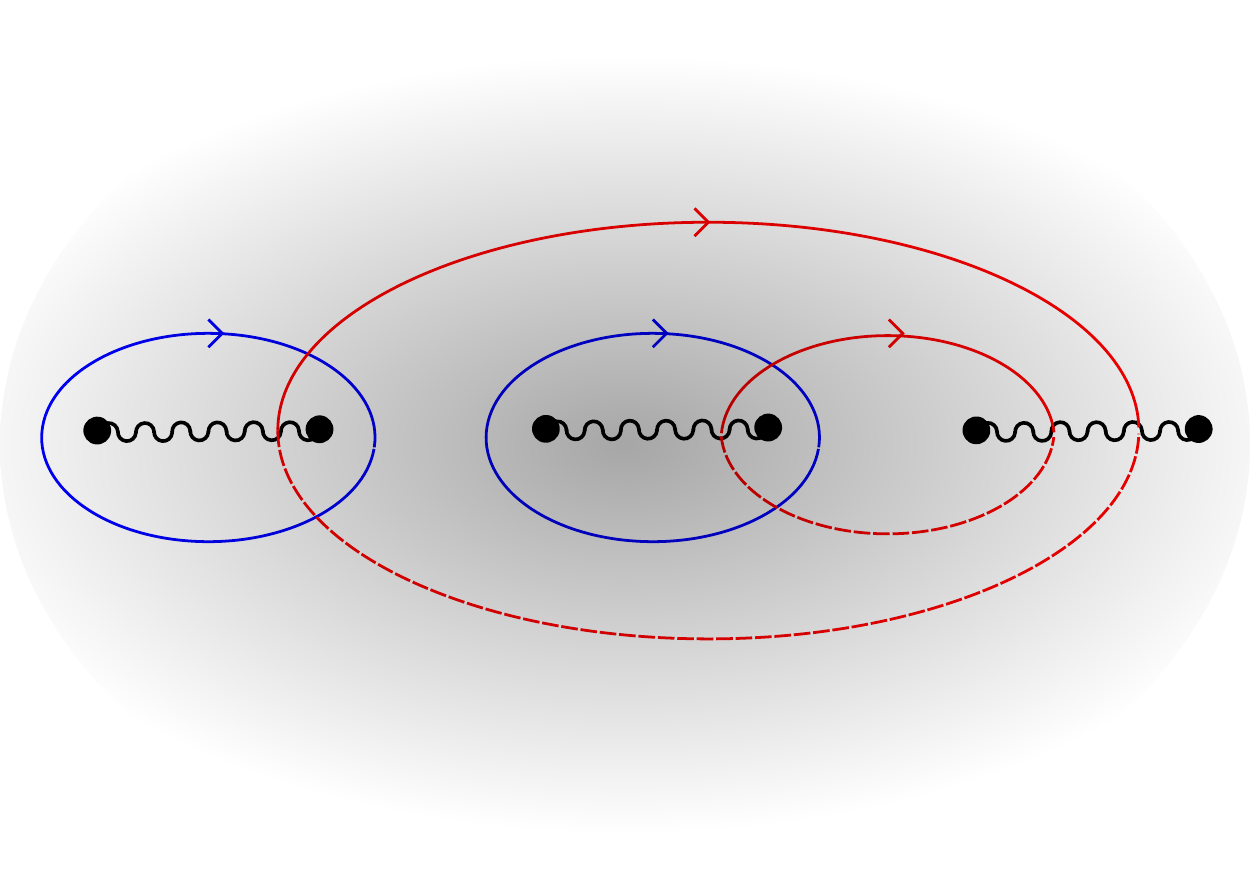}}%
    \put(0.05556548,0.32222531){\makebox(0,0)[lb]{\smash{$e_1$}}}%
    \put(0.23333809,0.32222531){\makebox(0,0)[lb]{\smash{$e_2$}}}%
    \put(0.41111071,0.32222531){\makebox(0,0)[lb]{\smash{$e_3$}}}%
    \put(0.59999411,0.32222531){\makebox(0,0)[lb]{\smash{$e_4$}}}%
    \put(0.76665594,0.32222531){\makebox(0,0)[lb]{\smash{$e_5$}}}%
    \put(0.92220698,0.32222531){\makebox(0,0)[lb]{\smash{$e_6$}}}%
    \put(0.07778705,0.44444399){\makebox(0,0)[lb]{\smash{$\gamma_1$}}}%
    \put(0.45555386,0.44444399){\makebox(0,0)[lb]{\smash{$\gamma_2$}}}%
    \put(0.48221975,0.54444108){\makebox(0,0)[lb]{\smash{${\tilde\gamma}_1$}}}%
    \put(0.66665884,0.46666556){\makebox(0,0)[lb]{\smash{${\tilde\gamma}_2$}}}%
  \end{picture}%
\endgroup%
    \end{center}
  \caption{A basis of cycles satisfying $\gamma_i \circ 
\widetilde\gamma_j=\delta_{ij}$ in the case $N=3$. The points $z_1,\cdots,z_6$ represent the six roots
of the SW curve for the SU(3) theory.}
  \label{fig:cycles}
\end{figure}
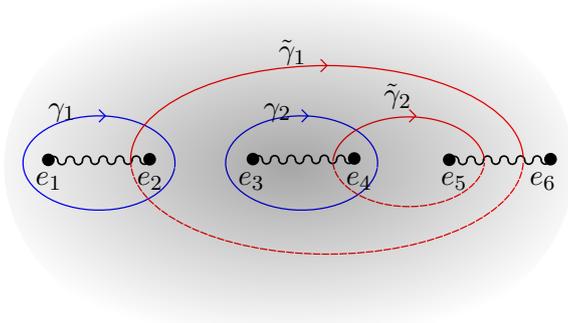

Let us now give some details that are useful for our purposes.
The curve (\ref{curve}), which was recently derived in Ref.~\cite{Fucito:2011pn} via a saddle point
analysis of the multi-instanton partition functions in the Nekrasov's approach,
defines a hyperelliptic fibration over a complex plane with 
a period matrix given by  
\begin{equation}
  \tau^{ij}=\frac{\partial {a_D}_i}{\partial a_j}
 \label{ta}
\end{equation}
where $a$ and $a_{D}$ are the periods
\begin{equation}
\begin{aligned}
a_i (e)=\frac{1}{2\ii\pi} \oint_{\gamma_i} \lambda~,~~~~
{a_D}_i(e) = \frac{1}{2  \ii\pi}\, \oint_{\widetilde \gamma_i} \lambda
\end{aligned}
\label{aad}
\end{equation}
of the SW 1-form differential $\lambda$ computed around a basis of cycles 
$(\gamma_i,\widetilde\gamma_j)$ on the $N$-cut complex plane
such that $\gamma_i \circ \widetilde\gamma_j=\delta_{ij}$. In Fig.~\ref{fig:cycles}
we have drawn an example of such cycles for the $N=3$ case.
The SW differential $\lambda$ can be written as
\begin{equation}
\lambda= z\, \Psi'(z)\, dz
\label{SWdifferentialnoi}
\end{equation}
with
\begin{equation}
\Psi(z) = \log \frac{P(z)+\sqrt{P^2(z) - g^2 Q(z)}}{\mu^{N}}~.
\label{psi}
\end{equation}
The chiral correlators $\big\langle\tr \phi^\ell\big\rangle$ of the SU($N$) theory 
on the D3$_0$ branes are then given by the integral
\begin{equation}
 \big\langle\tr\, \phi^\ell\big\rangle = 
\oint_{ \gamma} \frac{dw}{2  \ii\pi}~w^\ell\, \Psi'(w) 
\label{trj}
\end{equation}
with $\gamma$ a contour enclosing all $e$'s. Alternatively, they can be obtained
by expanding the generating functional
\begin{equation}
\Big\langle \tr \, \frac{1}{z-\phi } \Big\rangle  =  
  \Psi'(z) ~.
  \label{genphij}
\end{equation}
Integrating (\ref{genphij}) with respect to $z$, it is easy to find 
\begin{equation}
\Big\langle \tr\,\log \frac{z- \phi}{\mu}\Big\rangle 
=\log \frac{P(z)+\sqrt{ P(z)^2 - g^2 Q(z)}}{\mu^{N}} -\log\big(1+\sqrt{1-g^2}\big)
\label{trphi}
\end{equation}
where the integration constant has been fixed in order to match the $\cO(z^0)$ terms in the
expansion for large $z$ in both sides. 
With straightforward algebra and using (\ref{g2}), we can further rewrite 
our result in the following form
\begin{equation}
\Big\langle \tr\,\log \frac{z- \phi}{\mu}\Big\rangle 
=\frac{1}{2}\,\log \frac{P(z)+\sqrt{ P(z)^2 - g^2 Q(z)}}{P(z)-\sqrt{ P(z)^2 - g^2 Q(z)}}
+\frac{1}{2}\,\log \frac{Q(z)}{\mu^{2N}} + \frac{1}{2}\,\log q_0~.
\label{trphi1}
\end{equation}
Inserting (\ref{trphi1}) in (\ref{tauz}) and taking into account the explicit definition
of $Q$ given in (\ref{PQ}), we finally obtain
\begin{equation}
 \ii\pi  t (z)= \log
 \frac{P(z)-\sqrt{ P^2(z) - g^2 Q(z)}}{P(z)+\sqrt{P^2(z) - g^2 Q(z)}}~.
\label{tsugra}
\end{equation}
Our result exactly agrees with the one derived in Ref.~\cite{Cremonesi:2009hq} for the pure SU($N$) SYM theories using supergravity and M-theory 
considerations. Indeed, the pure SU($N$) theory with a dynamically generated scale $\Lambda$
can be obtained from the conformal one by decoupling all hypermultiplets through the following limit
\begin{equation}
 q_0\to 0~,~~~~m_u\to \infty\quad\mbox{with}\quad q_0\,\prod_{u=1}^Nm_u^{2}\equiv 
\Lambda^{2N}~~\mbox{fixed}~.
\label{limit}
\end{equation}
Using this prescription, (\ref{tsugra}) still holds with the replacement $g^2 Q(z) \to 4 \Lambda^{2N}$. 
This agrees with Eq.s~(2.13)-(2.15) of Ref.~\cite{Cremonesi:2009hq}. 
This solution can also be obtained without taking the decoupling limit of the conformal theory by directly starting with a configuration of $N$ D3 branes of type 0 and no D3 branes of type 1, in which 
case the $t$ profile computes the correlator (\ref{taupure}). 

Our solution (\ref{tsugra}) is also consistent with the results of Ref.~\cite{Witten:1997sc} 
where the SU($N)$ SYM theory is realized in type IIA using D4 branes stretched between two
NS branes. This set-up is related to our type IIB construction via a T-duality on the $S^1$ fiber 
of a two-centered Tab-Nut compactification of the $\mathbb C^2/\mathbb Z_2$ space. 
After T-duality the Tab-Nut space is mapped to a pair of NS5 branes at positions 
$\ln y_\pm$ on $S^1$ where 
 \begin{equation}
y_\pm(z)= P(z)\pm \sqrt{P^2(z) - g^2 Q(z)}
\end{equation}
and the twisted scalar $t$ parametrizes the distance between the two NS5 branes, namely
\begin{equation}
  \ii\pi t(z) = \log \frac{y_-(z) }{y_+(z)}
\label{ns5}
\end{equation}
in perfect agreement with (\ref{tsugra}). 
Finally we remark that the above formulae for the $t$ profile
are written in terms of the quantum moduli space coordinates $e$'s entering the SW geometry 
rather than in terms of the microscopic parameters $a$'s parametrizing the classical 
vacuum expectation values of the scalar fields. One can switch between the two descriptions
using the period formula (\ref{aad}) which relates the two sets of variables.

Let us now illustrate these results in the case of the SU(2) theory with four fundamental flavors.
In view of (\ref{sumem}), in this case we have $e_1=-e_2\equiv e$ and $m_1=-m_2\equiv m$.
The parameter $e$ is related to the classical vacuum expectation value $a$ of the
adjoint SU(2) scalar via the period formula (\ref{aad}), which up to 2-instantons reads
\begin{equation}
e= a-q_0\,\frac{(a^2-m^2)(3a^2+m^2)}{4 a^3} +q_0^2\,
\frac{(a^2-m^2)(27a^6+23 a^4 m^2+17 a^2 m^4-3 m^6)}{64 a^7} +\cdots~.
\label{ruleea}
\end{equation}
On the other hand, the SU(2) chiral  operators are encoded in the large $z$ 
expansion of (\ref{trphi}) for $N=2$, namely
\begin{eqnarray}
\Big\langle \tr_2\log \frac{z- \phi}{\mu}\Big\rangle \!&=& 
\log \frac{z^2}{\mu^2} -\sum_{\ell=1}^\infty\frac{\big\langle
\tr\,\phi^\ell\big\rangle}{\ell\,z^\ell}
= \log \frac{z^2}{\mu^2} -\frac{1}{2z^2}\left(\frac{2(1+q_0)}{1-q_0}e^2 -\frac{4q_0}{1-q_0}m^2\right)\label{chi2}
\\ &&\!\!\!\!-\frac{1}{4z^4}\left(\frac{2(1+q_0)^3}{(1-q_0)^3}e^4
 -\frac{16q_0(1+q_0)}{(1-q_0)^3}m^2e^2+\frac{4q_0(1+4q_0-q_0^2)}{(1-q_0)^3}m^4
\right)+\cdots ~.
\nonumber
\end{eqnarray}
Alternatively, one can exhibit the exact $z$ dependence of the left hand side of the above equation
as an expansion in $q_0$, whose first terms are
\begin{equation}
\Big\langle \tr_2\log \frac{z- \phi}{\mu}\Big\rangle =\log \frac{z^2-e^2}{\mu^2}
-q_0 \,\frac{(e^2-m^2)(2 z^2-e^2-m^2)}{(z^2-e^2)^2}+\ldots~.
\end{equation}
Using this expression and (\ref{ruleea}), one can write the profile of the twisted field
as a function of the microscopic parameters (that is $a$ and $m$) which up to 1-instanton is
\begin{equation}
 \begin{aligned}
  \ii\pi  t(z) &=
\ii\pi  t_0 
-2 \Big\langle \tr_2\log \frac{z- \phi}{\mu}\Big\rangle + 2 \log \frac{z^2-m^2}{\mu^2} \\
&= \ii\pi t_0 + 2 \log \frac{z^2-m^2}{z^2-a^2} + q_0\,
\frac{(a^2-m^2)^2\,(z^2+a^2)}{a^2(z^2-a^2)^2}+\cdots~.
 \end{aligned}
 \label{tam}
\end{equation}

We can therefore conclude that our methods provide a microscopic derivation of the 
supergravity profile for $t$ in which a direct relation with the chiral ring elements of the
gauge theory on the source branes is clearly established and the non-perturbative effects
are explicitly explained in terms of fractional D-instantons.

\subsection{The $t$ profile for the quiver gauge theory}

Now we consider the general case where the dynamics of both gauge groups of the quiver
theory is turned on. In this case the $t$ profile receives contributions from instantons 
of both types and the relevant correlators can be computed order by order 
in $q_0$ and $q_1$ with the help of localization techniques. In Appendix~\ref{MultiInstanton} we include a self-contained review of these techniques and a detailed derivation of the results. 
Here, as an example, we display the results for the case of the $\mathrm{SU}(2)\times \mathrm{SU}(2)$ quiver theory. 
For the first few instanton contributions one finds
\begin{equation}
 \begin{aligned}
\ii\pi t(z) =&\,
\ii\pi t_0 + 2 \log \frac{z^2-b^2}{z^2-a^2} +
(a^2-b^2)^2\left[q_0\frac{(z^2+a^2)}{a^2(z^2-a^2)^2}-q_1\frac{(z^2+b^2)}{b^2(z^2-b^2)^2}\right]\\
&-q_0\,q_1\frac{(a^2-b^2)^2(a^2+b^2)}{a^2b^2}\left[\frac{z^2+a^2}{(z^2-a^2)^2}
-\frac{z^2+b^2}{(z^2-b^2)^2}\right]+\cdots
 \end{aligned}
\label{quivert}
\end{equation}
where $a$ and $b$ are the vacuum expectation values of the scalar fields of the vector multiplets of the two SU(2) factors.
Note that the profile (\ref{quivert}) agrees with (\ref{tam}) after decoupling the 
$\mathrm{SU}(2)_1$ dynamics by sending $q_1\to 0$ and renaming $b\to m$.
Furthermore, (\ref{quivert}) provides a microscopic derivation of the first few 
D-instanton corrections to the gravity solution for the twisted scalar generated by a system of 
2 regular branes at the $\mathbb C^2/\mathbb Z_2$ singularity.

\section{The relation between $t$ and effective gauge couplings}
\label{secn:gaugegrav}

We now provide an interpretation of the profile function $t(z)$ generated by a symmetric 
D3 brane source as the effective IR coupling constant of a SU($N$) theory with a 
Coulomb branch parametrized by the coordinate $z$. 
In particular, we will find that the non-perturbative
contributions in the $t$ solution spoil the simple correspondence (\ref{taut})
and lead to a more sophisticated relation between the twisted scalar $t$ and the gauge
coupling.

The solution (\ref{tsugra}) exhibits an explicit dependence on the SU($N$) mass invariants
$T_\ell$ and on the SU($N$) gauge invariants $u_\ell$, defined in (\ref{PQ}),
which parametrize the quantum moduli space of the SU($N$) source theory. The analysis at a
generic point in moduli space is left to future investigations.
Here we focus on a particular point, the so-called ``special vacuum'' \cite{Argyres:1999ty}, 
where the gauge and flavor invariants are taken to be
\begin{equation}
\begin{aligned}
& T_\ell= u_\ell=0~\quad\mbox{for}~~\ell=1,\cdots,N-1~,\\
& T_N=-m^N~,~~~u_N=-e^N~.
\end{aligned}
\label{sp}
\end{equation}
At this point the U(1)$_R$ symmetry is partially restored to 
a ${\mathbf{Z}}_N$ symmetry and the Coulomb branch is described
by a single parameter. In our brane construction the special vacuum is realized by placing 
the fractional D3 branes in a symmetric configuration around the origin of the transverse space, 
namely by taking the positions $a_u$ and $m_u$ of the D3$_0$ and D3$_1$ branes respectively, as follows
\begin{equation}
 a_u= a \,\omega^{u-1}~,~~~~m_u= m \,\omega^{u-1}
\label{simam}
\end{equation}
with $u=1,\cdots, N$, and $\omega= \ee^{ 2\ii\pi /N}$.
Exploiting the relations (\ref{aad}), one can prove that this choice implies that 
also the quantum variables $e_u$ are symmetrically
distributed around the origin:
\begin{equation}
 e_u= e \,\omega^{u-1}
\label{ei}
\end{equation}
which implies (\ref{sp}). 
Moreover, one can check that
\begin{equation}
 \langle\, \tr \,\phi^\ell\,\rangle=0~\quad\mbox{for}~~\ell=1,\cdots,N-1
\label{trf}
\end{equation}
and that the relation between the quantum expectation value of order $N$
\begin{equation}
{\mathbf v} \equiv \frac{1}{N}\,\langle\, \tr\, \phi^N\,\rangle
\label{vv0}
\end{equation}
and the parameter $u_N$ of the SW curve is
\begin{equation}
 u_N = -{\mathbf v} - \frac{2q_0}{1+q_0}\,\big(m^N -{\mathbf v} \big)
 = -\sqrt{1-g^2}\,\vv - \frac{2q_0}{1+q_0}\,m^N~.
\label{vv}
\end{equation}
The expectation value $\mathbf v$ is the appropriate variable to describe the quantum moduli
space of the theory at the special vacuum and can be regarded as the quantum equivalent
of the classical variable parametrizing the transverse space of the fractional D3 branes.
Just like in the perturbative analysis \cite{Bertolini:2000dk,Polchinski:2000mx,Billo:2001vg,Bertolini:2001qa}
one puts all source branes at the origin of the transverse space, also here 
we do a similar choice and take ${\mathbf v}=0$.
Notice that this corresponds to classically arranging the branes in a sort of an
extended configuration resembling that of the enhan\c{c}on ring and for this reason
the $\vv=0$ vacuum is sometimes called the ``enhan\c{c}on vacuum'' \cite{Benini:2008ir,Cremonesi:2009hq}.
For this choice of parameters the characteristic polynomials entering the SW curve (\ref{curve}) become
\begin{equation}
 P_{\mathrm{enh}}(z)=\left( {z}^N-\frac{2q_0}{1+q_0}\,{m}^N\right)~,
\qquad Q_{\mathrm{enh}}(z)=\big(z^N-m^N\big)^2~,
\label{curve1}
\end{equation}
and the twisted scalar emitted by the branes in the enhan\c{c}on vacuum is 
 \begin{equation}
\begin{aligned}
  \ii\pi t_{\mathrm{enh}}(z) &= \log
 \frac{P_{\mathrm{enh}}(z)-\sqrt{ P_{\mathrm{enh}}^2(z) - g^2 Q_{\mathrm{enh}}(z) }}
{P_{\mathrm{enh}}(z)+\sqrt{P_{\mathrm{enh}}^2(z) -g^2 Q_{\mathrm{enh}}(z)}}
~.
\label{tsugrapure}
\end{aligned}
\end{equation}
For later convenience it is useful also to write the exponentiated form of the solution 
as a power series obtained by expanding (\ref{tsugrapure}) for small values of $q_0$
\begin{equation}
q_{\mathrm{enh}}(z) \equiv 
\ee^{\ii\pi t_{\mathrm{enh}}(z)}
= q_0\Big(1-\frac{m^N}{z^N}\Big)^2\Big[1+q_0\frac{2m^{2N}}{z^{2N}}
+q_0^2\frac{m^{2N}}{z^{2N}}\Big(2-\frac{4m^N}{z^N}+\frac{5m^{2N}}{z^{2N}}\Big)+
\cdots\Big]~.
\label{qmz}
\end{equation}
As is clear from this explicit expression, the field $q_{\mathrm{enh}}$ depends only on $q_0$ 
and on the ratio $\frac{m^N}{z^N}$. 
To emphasize this feature, from now on we will denote the enhan\c{c}on profile as $q_{\mathrm{enh}}(\frac{m^N}{z^N},q_0)$, instead of simply $q_{\mathrm{enh}}(z)$.
If we set $m=0$, the theory is conformal and $q_{\mathrm{enh}}$ becomes constant:
\begin{equation}
q_{\mathrm{enh}}(0,q_0) =q_0~.
\label{tmzero}
\end{equation}

In the following, we will show that $q_{\rm enh}$ encodes the information about the gauge coupling
of the SU$(N)$ theory with $2N$ massive flavors at the special vacuum. The chiral dynamics of 
this theory is described by the SW curve
\begin{equation}
\label{massive1}
y^2 = (x^N + u_N)^2 - g^2 (x^N - m^N)^2
\end{equation}
which follows from specializing (\ref{curve}) with the simple form that the $P$ and $Q$ polynomials
assume in the special vacuum%
\footnote{We have employed the complex coordinates $(y,x)$ rather than $(y,z)$ as in (\ref{curve}) since the variable $z$ plays, in the following discussion, the r\^ole of a scale.}. 
This curve can actually be obtained from the massless SW curve%
\footnote{In the massless case, after taking into account (\ref{vv}) for $m=0$, 
the curve (\ref{curve}) becomes: $ y^2=(z^N-\sqrt{1-g^2}\vv)^2-g^2z^{2N}$. The dependence
on $\vv$ is actually fictitious and can be absorbed
by the rescalings $z^N\to -\sqrt{1-g^2}\,\vv\,X^N$, $y^2 \to (1-g^2)\,\vv^2 Y^2$, which
bring the curve in the form (\ref{conf1}).} 
\begin{equation}
\label{conf1}
Y^2 = (X^N +1)^2 - g^2 \,X^{2N}
\end{equation}
after the replacement
\begin{equation}
\label{q0toqe}
q_0 ~\longrightarrow ~ \qen\left(\frac{m^N}{\vv},q_0\right)~.
\end{equation}
Indeed, from (\ref{vv}) and (\ref{tsugrapure}) one finds that 
(\ref{q0toqe}) implies
\begin{equation}
g^2=\frac{4q_0}{(1+q_0)^2} ~\longrightarrow ~  \frac{4\qen}{(1+\qen)^2}
=g^2\,\frac{Q_\mathrm{enh}}{P^2_\mathrm{enh} }= g^2\,  \frac{(u_N+m^N)^2}{(u_N + g^2\, m^N)^2}~.
\end{equation}
Plugging this result in (\ref{conf1}), it is easy to show that the massless curve 
takes exactly the form (\ref{massive1}) of the massive one if we perform the change of variables  
\begin{equation}
\label{change1}
X^N = \frac{u_N + g^2\, m^N}{1-g^2}\,\frac{1}{x^N}~,~~~~
\frac{Y^2}{X^{2N}}= \frac{1 - g^2}{(u_N + g^2\, m^N)^2}\,y^2~.
\end{equation}
The above derivation can be readily adapted to the case of a pure SU$(N)$ theory at the special vacuum, to show that it can be equivalently represented as a conformal theory with an UV coupling given by the twisted field dual to the enhan\c{c}on vacuum of that theory.

Summarizing the SW curve of the massive theory with UV coupling $q_0$ at the special vacuum 
coincides with the curve for a SU$(N)$ theory with $2N$ \emph{massless} flavors and UV coupling
given by $\qen$ evaluated at $z=\vv^{1/N}$.
Since SW curves encode the low-energy dynamics, the equivalence shown above implies
in particular that the effective coupling constants $\tau^{ij}\big(\frac{m^N}{\mathbf v},q_0\big)$ 
of the massive SU($N$) theory in the special vacuum can be expressed as the IR couplings 
of a conformal SU($N$) theory with a UV coupling given by $q_{\mathrm{enh}}\Big(\frac{m^N}{\mathbf v},q_0\Big)$, namely
\begin{equation}
 \tau^{ij}\left(\frac{m^N}{\mathbf v},q_0\right)=
 \tau^{ij}\left(0,q_{\mathrm{enh}}\Big(\frac{m^N}{\mathbf v},q_0\Big)\right)~.
\label{tauandt}
\end{equation}
We will now verify the relation (\ref{tauandt}) for SU($N$) theories with $N=2,3,4$, 
deriving and exploiting the explicit form of the corresponding conformal ``UV/IR'' relation
$\tau^{ij}(0,q_0)$.

\subsection{SU(2) theory}
\label{secn:su2}

The first example we consider is the SU(2) theory with 4 fundamental flavors.
In this case we have a single effective coupling constant, $\tau_{\mathrm{SU(2)}}\equiv \tau$
which is a function of the flavor masses and the tree-level coupling $t_0$.
Through explicit multi-instanton computations (see Appendix \ref{MultiInstanton} for some details)
one finds that, when the flavors are massless, $\tau$ is
related to $q_0$ as follows 
\begin{equation}
\ii\pi \tau(0,q_0)\,\equiv \,\ii\pi \tau =\log q_0 +\ii \pi -\log 16+\frac{1}{2}\,q_0+\frac{13}{64}\,q_0^2
+\frac{23}{192}\,q_0^3+\cdots~.
\label{taq0m}
\end{equation}
It is interesting to observe that the inverse relation can be expressed in terms of modular functions. Indeed, inverting (\ref{taq0m}) we obtain
\begin{equation}
q_0=-16 \,\big( \ee^{\ii\pi\tau} +8 \,\ee^{2\ii\pi \tau}+44\, \ee^{3\ii\pi \tau}+ \cdots\big) = -16\,\frac{\eta^8(4\tau)}{\eta^8(\tau)}
\label{q0t1}
\end{equation}
where $\eta(\tau)$ is the Dedekind $\eta$-function%
\footnote{In our conventions
the Dedekind $\eta$-function is defined as
$\eta(\tau)= \ee^{\frac{\ii\pi \tau}{24}} \prod_{n=1}^\infty (1-\ee^{\ii\pi n \tau})$.
Notice that $t_0$ can be alternatively written in the more standard elliptic form: 
$q_0=-\frac{\theta_2^4(\tau)}{\theta_4^4(\tau)}$ which shows that 
$\ee^{\ii\pi t_0}$ can be thought as one of the harmonic ratios 
of an elliptic curve. Our conventions for the $\theta$-functions are the same as in Ref.~\cite{Billo:2010mg}.}. This relation expresses the UV coupling $q_0$ in terms of the IR one $\tau$. 

On the other hand, if in (\ref{taq0m}) we replace $q_0$ with the SU(2) enhan\c{c}on solution given by (\ref{qmz}) with $N=2$, after some simple algebra we find 
\begin{equation}
\begin{aligned}
\ii\pi\tau\left(0,q_{\mathrm{enh}}\Big(\frac{m^2}{z^2},q_0\Big)\right)
&=\log q_{\rm enh} +\ii \pi -\log 16+\frac{1}{2}\,q_{\rm enh} +\frac{13}{64}\,q_{\rm enh} ^2
+\cdots\\
&=\log q_0+\ii \pi  -\log 16+2\log\Big(1-\frac{m^2}{z^2}\Big)
+\,q_0\Big(\frac{1}{2}- \frac{m^2}{z^2} +  \frac{5m^4}{2z^4}\Big)\\
&~~~~+\,q_0^2\Big(\frac{13}{64}-\frac{13m^2}{16 z^2}+\frac{135m^4}{32z^4}
-\frac{109m^6}{16z^6}+\frac{269m^8}{64z^8}\Big)+\cdots~.
\end{aligned}
\label{taq0}
\end{equation}
This expression exactly agrees with the running coupling of a $\mathcal N=2$ SU($2$) gauge theory
with 4 hypermultiplets of masses $(m,-m,m,-m)$ that can be deduced from 
the formulae of Appendix~\ref{MultiInstanton} (see in particular the subsection~\ref{sunspecial}), 
provided we trade $z^2$ for the quadratic invariant $\mathbf v$ parametrizing the Coulomb branch of the special vacuum, namely $z^2~\leftrightarrow~{\mathbf v}$.
In other words we have shown that
\begin{equation}
\tau\left(\frac{m^2}{\mathbf v},q_0\right)=
 \tau\left(0,q_{\mathrm{enh}}\Big(\frac{m^2}{\mathbf v},q_0\Big)\right)
\label{tauandt2}
\end{equation}
which is precisely the relation (\ref{tauandt}) for SU(2).

We point out that the relation (\ref{tauandt2}) can actually be proved in full generality 
by exploiting the SW curve (\ref{curve1}) with $N=2$ 
(see Appendix \ref{secn:appc} for some details), and that 
our results can be readily extended to the case of the pure SU(2) theory by taking the 
decoupling limit (\ref{limit}).
In this case, as one can see from (\ref{taq0}), the non-perturbative part of $\tau$ becomes
\begin{equation}
 \ii\pi\tau\big|_{\mathrm{n.p.}}=
\frac{5\Lambda^4}{2{\mathbf v}^2}+\frac{269\Lambda^8}{64{\mathbf v}^4}+\cdots
\label{puresu2inst}
\end{equation}
in perfect agreement with the instanton calculations%
\footnote{Usually the instanton contribution to the pure SU(2) coupling constant 
is written in terms of the classical parameter 
$a^2=\frac{1}{2}\tr\langle \phi^2\rangle$, as $\ii\pi\tau|_{\mathrm{n.p.}}=
\frac{3\Lambda^4}{2a^4}+\frac{105\Lambda^8}{64a^8}+\cdots$. Taking into account the relation between $a^2$ and $\mathbf v$, one can check that this is equivalent to (\ref{puresu2inst}).}.
We stress that the case of the pure SYM theory can be treated from the very beginning without making reference to the conformal theory and its decoupling limit if one considers a stack of 2 fractional D3 branes of type 0 and no branes of type 1.
In this case one starts again from (\ref{taq0m}) and replace $q_0 $ with $q_{\rm enh}$ given by 
(\ref{tsugrapure}) with $g^2 Q(z)=4 \Lambda^4$. 
The results agree with the SU(2) effective coupling given in (\ref{puresu2inst}).

\subsection{SU(3) theory}
\label{secn:su3}

Let us now consider the SU(3) theory with 6 fundamental flavors. In the special vacuum, 
the matrix of coupling constants has the classical form even when the
1-loop and instanton corrections are taken into account, namely
\begin{equation}
2\pi\ii\,\tau^{ij}_{\mathrm{SU}(3)}= \begin{pmatrix}
     \,2&~ 1 \,\\
     \,1& ~2 \,\\
     \end{pmatrix}\,\pi\ii\,\tau
\label{tautree3}
\end{equation}
where the effective coupling $\tau$ is a function of the flavor masses and the UV coupling
$q_0$.
When the flavor masses are zero, explicit 1-loop and multi-instanton calculations allow to
establish that
\begin{equation}
\ii\pi\tau(0,q_0)\,\equiv \,\ii\pi\tau = \log q_0 +\ii \pi -\log 27+\frac{4}{9}\,q_0+\frac{14}{81}\,q_0^2
+\frac{1948}{19683}\,q_0^3+\cdots
\label{taq0m3}
\end{equation}
from which one derives
\begin{equation}
q_0=-27 \,\big( \ee^{\ii\pi\tau} +12 \,\ee^{2\ii\pi\tau}+
90\, \ee^{3\ii\pi\tau}+ \cdots\big) =-27\,\frac{\eta^{12}(3\tau)}{\eta^{12}(\tau)}~.
\label{q0t3}
\end{equation}
This relation can be proved in full generality by considering the SW curve discussed in
Ref.~\cite{Minahan:1995er} and comparing it to the one in (\ref{curve1}) (some details are given in
Appendix~\ref{secn:appc}).

Without any further ado, as in the previous case we extend the massless relations 
to the non-conformal case by simply replacing $q_0$ with 
$q_{\mathrm{enh}}(\frac{m^3}{\mathbf v},q_0)$.
In this way we obtain
\begin{equation}
\begin{aligned}
\ii\pi\tau\left(0,q_{\mathrm{enh}}\Big(\frac{m^3}{\mathbf v},q_0\Big)\right)&=
\log q_{\mathrm{enh}} +\ii \pi -\log 27+\frac{4}{9}\,q_{\mathrm{enh}}+\frac{14}{81}\,q_{\mathrm{enh}}^2
+\cdots\\
&=\log q_0+\ii \pi  -\log 27 +2\log\Big(1-\frac{m^3}{{\mathbf v} }\Big)
+\,q_0\Big(\frac{4}{9}- \frac{8m^3}{9{\mathbf v}} +  \frac{22m^6}{9{\mathbf v}^2}\Big)\\
&~~~~+\,q_0^2\Big(\frac{14}{81}-\frac{56m^3}{81{\mathbf v}}+\frac{106m^6}{27{\mathbf v}^2}
-\frac{524m^9}{81{\mathbf v}^3}+\frac{329m^{12}}{81{\mathbf v}^4}\Big)+\cdots~.
\end{aligned}
\label{taq03}
\end{equation}
One can check that this expression exactly agrees with the running coupling 
of a $\mathcal N=2$ SU(3) gauge theory
with 6 hypermultiplets at the special vacuum parametrized by 
the cubic invariant $\mathbf v=\frac{1}{3}\langle\tr \,\Phi^3\rangle$
(see in particular the subsection \ref{sunspecial}). Thus, also for the SU(3) theory
the relation (\ref{tauandt}) is verified.

Again, the pure SU(3) theory can be obtained by decoupling the 6 flavors or by repeating the entire derivation starting from a microscopic configuration with only 3 fractional D3 branes of type 0. 

\subsection{SU(4) theory}
\label{secn:su4}
The last explicit example we consider is the SU(4) theory with 8 fundamental flavors. As observed
in Ref.s~\cite{Minahan:1996ws,Aharony:1996ks}, the 1-loop and instanton corrections  spoil 
the classical form of the coupling constant matrix even if one works in
the special vacuum. In fact, in the massless case two different matrix
structures appear:
\begin{equation}
2\pi\ii\,\tau^{ij}_{\mathrm{SU}(4)}= \begin{pmatrix}
     \,2&~ 1&~1 \,\\
     \,1& ~2&~1\,\\
     \,1& ~1&~2\,\\
     \end{pmatrix}\pi\ii\,\tau \,+\,
     \begin{pmatrix}
     0&-1&1\\
     -1&-2&-1\\
     1&-1& 0\\
     \end{pmatrix}\pi\ii\,\tau'
\label{tautree4}
\end{equation}
where the two couplings $\tau$ and $\tau'$ are given by
\begin{equation}
\begin{aligned}
\ii\pi\tau_+ \equiv \,\ii\pi\big(\tau+ \tau'\big) &= \log q_0 + \ii \pi -\log 16 +
\frac{1}{2}\,q_0+\frac{13}{64}\,q_0^2 +\frac{23}{192}\,q_0^3+\cdots~. \\
\ii\pi\tau_- \equiv \,\ii\pi\big(\tau - \tau'\big) &=
\log q_0 + \ii \pi -\log 64 +\frac{3}{8}\,q_0+\frac{141}{1024}\,q_0^2
+\frac{311}{4096}\,q_0^3+\cdots~.
\end{aligned}
\label{taq0m4}
\end{equation}
These relations can be inverted yielding
\begin{equation}
\begin{aligned}
q_0 &=-16 \,\big( \ee^{\ii\pi\tau_+} +8 \,\ee^{2\ii\pi\tau_+}+
44\, \ee^{3\ii\pi\tau_+}+ \cdots\big) =
-16\,\frac{\eta^8(4\tau_+)}{\eta^8(\tau_+)}~,\\
q_0 &=-64 \,\big( \ee^{\ii\pi\tau_-} +24 \,\ee^{2\ii\pi\tau_-}+
300\, \ee^{3\ii\pi\tau_-}+ \cdots\big) =
-64\,\frac{\eta^{24}(2\tau_-)}{\eta^{24}(\tau_-)}~.
\end{aligned}
\label{q0t4}
\end{equation}
It is interesting to observe that also in the SU(4) theory one can write exact formulae in terms
of ratios of Dedekind $\eta$-functions. By replacing in the right hand side of (\ref{taq0m4}) $q_0$
with the $q_{\mathrm{enh}}(\frac{m^4}{\mathbf v},q_0)$, we obtain the coupling constants
for the massive SU(4) theory. After simple algebra, we find
\begin{eqnarray}
\ii\pi\tau_+\left(0,q_{\mathrm{enh}}\Big(\frac{m^3}{\mathbf v},q_0\Big)\right)
 &=&\log q_{\mathrm{enh}} +\ii \pi -\log 
16+\frac{1}{2}\,q_{\mathrm{enh}}+\frac{13}{64}\,q_{\mathrm{enh}}^2
+\cdots\label{su4p}\\
&=&\log q_0+\ii \pi  -\log 16+2\log\Big(1-\frac{m^4}{ \mathbf{v} }\Big)
+\,q_0\Big(\frac{1}{2}- \frac{m^4}{\mathbf{v}} +  \frac{5m^8}{2\mathbf{v}^2}\Big)\nonumber\\
&&\!\!+\,q_0^2\Big(\frac{13}{64}-\frac{13m^4}{16\mathbf{v}}+\frac{135m^8}{32 \mathbf{v}^2}
-\frac{109m^{12}}{16\mathbf{v}^3}+\frac{269m^{16}}{64\mathbf{v}^4}\Big)+\cdots
\nonumber
\end{eqnarray}
and
\begin{eqnarray}
\ii\pi\tau_-\left(0,q_{\mathrm{enh}}\Big(\frac{m^3}{\mathbf v},q_0\Big)\right)&=&\log q_{\mathrm{enh}} +\ii \pi -\log 64+\frac{3}{8}\,q_{\mathrm{enh}}+\frac{141}{1024}\,q_{\mathrm{enh}}^2
+\cdots\label{su4m}\\
&=&\log q_0+\ii \pi  -\log 64+2\log\Big(1-\frac{m^4}{\mathbf{v} }\Big)
+\,q_0\Big(\frac{3}{8}- \frac{3m^4}{4z^4} +  \frac{19m^8}{8\mathbf{v}^2 }\Big)\nonumber\\
&&\!\!+\,q_0^2\Big(\frac{141}{1024}-\frac{141m^4}{256\mathbf{v} }+\frac{1831m^8}{512 \mathbf{v}^2 }
-\frac{1549m^{12}}{256\mathbf{v}^3}+\frac{3981m^{16}}{1024\mathbf{v}^4}\Big)+\cdots~.
\nonumber
\end{eqnarray}
Again one can check that these expressions agree with the running couplings of the SU(4) theory
with 8 massive flavors in the special vacuum parametrized by  the quartic
invariant $\mathbf v=\frac{1}{4}\langle\tr \,\Phi^4\rangle$ (see for example the subsection~\ref{sunspecial}). Thus, also for the SU(4) theory
the relation (\ref{tauandt}) is verified.

\section{Summary of results and conclusions}
\label{secn:summary}
We can summarize our results in the following three main points.
\paragraph{$\bullet$}First, we have provided an explicit derivation of the supergravity profile
for the twisted scalar $t$ sourced by a configuration of fractional branes in the 
$\mathbb C^2/\mathbb Z_2$ orbifold by computing the emission diagrams at the disk level. 
Technically, these diagrams involve one closed and many open string insertions on disks
ending on fractional D3 or D(--1) branes.
Disks with the boundary on the D3 branes are responsible for the logarithmic 
profile of $t$ which is typical of perturbation theory, whereas 
disks with the boundary conditions of fractional D(--1) branes generate
the non-perturbative corrections associated to instantons. This analysis is therefore
a microscopic explanation of the complete supergravity
solution for $t$ in terms of source branes. As a result of this approach we are able to write $t$ in terms of the quantum observables (chiral correlators) of the quiver gauge theory defined on the source 
D3 branes (see (\ref{tauz})).  This relation provides a clear example of a gauge/gravity correspondence 
in which a (six-dimensional) supergravity field is expressed as a quantum correlator of 
a (four-dimensional) gauge theory.

\paragraph{$\bullet$} Second, we have shown how to compute the non-perturbative instanton contributions 
in the $t$ profile using localization techniques and Nekrasov's integrals, order by order in the
instanton number. We have also explained how to derive these non-perturbative effects from the SW curve associated to the gauge theory on the source branes, a procedure which essentially resums the instanton expansion. In this way the exact supergravity $t$ profile can be written in a very compact and simple,
but also very explicit, form (see (\ref{tsugra})).

\paragraph{$\bullet$} Third, we have investigated the relation between the supergravity solution
for $t$ and the effective gauge couplings of SU($N$) theories. At tree-level, by considering
the world-volume action of $N$ fractional D3 branes, one can easily show that $t$ plays the
r\^ole of the complexified coupling constant of the SU($N$) gauge theory. However, 1-loop and instanton
corrections spoil this simple identification. The profile $t(z)$ is related to a chiral correlator rather than a coupling 
in the gauge theory.  Nevertheless, our results show that the
$t$ profile still encodes a lot of information about the effective gauge couplings in a rather
non-trivial way. In particular, the field $t_{\mathrm{enh}}(z)$ emitted by
a configuration of branes sitting at the origin of the quantum moduli space (see (\ref{tsugrapure})), 
turns out to be related to the effective gauge coupling of SU($N$) theories in the special vacuum through modular functions linking the UV and IR couplings of the massless theory (see (\ref{q0t1})). Moreover, the transverse coordinate $z$ at which the supergravity solution is evaluated is identified with the coordinate $\mathbf v$ parametrizing the quantum moduli space of the SU($N$) theory in the special vacuum according to $z^N~\leftrightarrow~{\mathbf v}~.$

\paragraph{}~
It would be interesting to extend the analysis presented here in several ways. In particular it
would be nice to obtain the supergravity $t$ profile in a compact form also for the quiver
theory, namely when both nodes of the quiver diagram are treated dynamically. It would be
also interesting to investigate the relation between $t(z)$ and $\tau^{ij}$ 
away from the special vacuum. Moreover, we think that our results could be useful in studying the
strong coupling regime of the superconformal $\mathcal N=2$ gauge theories from a dual perspective
and providing the basis for the studies of ${\cal N}=1$ settings. 
We hope to return to some of these issues in the near future.

\vskip 1cm
\noindent {\large {\bf Acknowledgments}}
\vskip 0.2cm
The authors would like to thank S. Cremonesi, L. Martucci, I. Pesando and R. Russo for several very useful discussions.
\vskip 1cm
\appendix

 \section{ Multi-instanton computations and Nekrasov's integrals}
\label{MultiInstanton}

In this appendix we review the computation via localization techniques of multi-instanton corrections 
to the prepotential and chiral correlators in ${\cal N}=2$ SYM theories.
Although we are mainly interested in SU($N_0$)$\times$SU($N_1$) quiver theories, 
for future convenience we present a self-contained description of other 
interesting models. We also consider the more general case of U($N$) gauge symmetry groups
rather than the special case of SU($N$), to which one can always reduce by imposing the
appropriate constraints.

The dynamics of $\mathcal N=4$ U($N$) gauge theories in four dimensions can be described in terms
of open strings ending on a stack of $N$ D3 branes in flat space-time.
The $\mathcal N=4$ supersymmetry can be broken to $\mathcal N=2$ by a mass deformation or by 
an orbifold projection. 
In the first case the resulting gauge theory contains a massless $\mathcal N=2$ vector
multiplet and a massive hypermultiplet, both in the adjoint representation of SU($N$). 
Decoupling the massive multiplet by sending its mass to infinity, one obtains the pure
SU($N$) SYM theory, and by adding D7 branes one can introduce fundamental matter.
In the orbifold case, instead, one finds in general a quiver gauge theory with bi-fundamental
matter.

The mass deformation and/or the orbifold projection as well as the brane
arrangement break the isometry group of the flat ten-dimensional space-time
(with euclidean signature) according to
\begin{equation}
 \mathrm{SO}(10)~\rightarrow~\Big(\mathrm{SU}(2)_L\times \mathrm{SU}(2)_R\Big)\times 
\Big(\widehat{\mathrm{SU}(2)}_L\times \widehat{\mathrm{SU}(2)}_R\Big)\times \mathrm{SO}(2)~.
\label{lorentz}
\end{equation}
In the following we find it convenient to label the representations of this group by the 
spinor indices $(\alpha,\dot{\alpha})$, $(a,\dot{a})$ and
$(+,-)$ of the various SU(2) and SO(2) factors, and parametrize the Cartan of the SU(2)$^4$
subgroup in terms of four parameters $(\epsilon_1,\epsilon_2,\epsilon_3,\epsilon_4)$.
The spinor weights will then be written as $\frac{1}{2}\big(\pm\epsilon_1\pm\epsilon_2\big)$,
$\frac{1}{2}\big(\pm\epsilon_3\pm\epsilon_4\big)$ with an even (odd) number of pluses for
undotted (dotted) indices respectively.

\subsection{D-instanton moduli}

The moduli space of instanton solutions in $\mathcal N=2$ gauge theories with topological charge $k$
can be realized by introducing $k$ D(--1) branes and considering all possible open strings with at
least one end-point on them. The essential instrument to compute the corresponding
multi-instanton partition function is the 
localization procedure based on the cohomological structure of the moduli action 
which is exact with respect to a suitable BRST charge $Q$:
\begin{equation} 
S_{\mathrm{inst}}= Q \,\Xi~.
\label{q}
\end{equation}
$Q$ is determined by selecting one component of the conserved supersymmetry charges.
This choice breaks the SU(2)$^4$ subgroup of the Lorentz group (\ref{lorentz})
to the $\mathrm{SU}(2)^3$ subgroup which leaves $Q$ invariant. In our case we take this 
$\mathrm{SU}(2)^3$ subgroup to be given by
\begin{equation}
\mathrm{SU}(2)_1\times \mathrm{SU}(2)_2\times \mathrm{SU}(2)_3=\mathrm{SU}(2)_L\times\widehat{\mathrm{SU}(2)}_L\times\diag\big[\mathrm{SU}(2)_R
\times\widehat{\mathrm{SU}(2)}_R\big]~,
\end{equation}
which in practice means identifying the spinor indices $\dot{\alpha}$ and $\dot{a}$.
After this twist, the four $\epsilon$-parameters become subject to the restriction
\begin{equation}
 \epsilon_1+\epsilon_2+\epsilon_3+\epsilon_4=0~.
\end{equation}
We now list the open string modes corresponding to physical excitations of open strings with
at least one end-point on the D-instantons. We distinguish the neutral sector, {\it i.e.} the 
D(--1)/D(--1) strings, the charged sector, {\it i.e.} the D(--1)/D3 or D3/D(--1) strings, and
if also D7 branes are present, the flavored sector with the D(--1)/D7 or D7/D(--1) strings.

\paragraph{Neutral sector:} The physical modes of the D(--1)/D(--1) open strings comprise:
two complex bosons~\footnote{We write bispinors as two complex fields given in terms of
the two complexified gamma matrices
 $\widehat \sigma_1=\widehat \sigma_3=\sigma_3+i \sigma_0$ and 
$\widehat\sigma_2=\widehat \sigma_4=\sigma_1+\ii \sigma_2$.} 
$B_\ell=\widehat \sigma^{\alpha\dot{\alpha}}_\ell B_{\alpha\dot\alpha}$ (with $\ell=1,2$) 
from the NS sector which parametrize the positions of the instantons in spacetime,
plus the corresponding fermionic partners $M_\ell=\widehat \sigma^{\alpha\dot{a}}_\ell   
M_{\alpha\dot{a}}$, $\eta=\epsilon^{\dot{\alpha}\dot{a}}\lambda_{\dot{\alpha}\dot{a}}$ and
$\lambda_c=\sigma_{c}^{\dot{\alpha}\dot{a} } \lambda_{\dot{\alpha}\dot{a}}$ (with
$c=1,2,3$) from the R sector. Furthermore, to have a linear SUSY transformation for these 
latter fermions, we must add three auxiliary fields $D_c$, while $\eta$ is paired with the complex bosonic field $\bar \chi$ parametrizing the positions of the instantons in the transverse space.
We then have two other complex bosons $B_{\dot\ell}= \widehat \sigma^{a\dot{a}}_{\dot \ell}
B_{a\dot a}$ (with $\dot\ell=3,4$) from the NS sector and their fermionic partners from the R sector, namely $M_{\dot \ell}=
\widehat \sigma^{a\dot{\alpha} }_{\dot \ell}   M_{a \dot{\alpha} }$
and $\lambda_{m'}=
\hat \sigma^{\alpha a}_{m'}  \lambda_{\alpha a}$ (with $m'=3,4$). 
In analogy with what we did before, to have a linearly realized SUSY 
we add two complex auxiliary fields $D_{m'}$.
Finally we have another complex field $\chi$ which is a SUSY singlet.
All these neutral moduli transform in the adjoint representation of U($k$).

\paragraph{Charged sector:} The physical modes of the D(--1)/D3 open strings 
provide two complex bosonic fields $w_{\dot \alpha}$ from the NS sector, and their fermionic superpartners $\mu_{\dot a}$ from the R sector. Here, we have two extra fermionic zero-modes
$\mu_a$ which are paired with the auxiliary fields $h_a$. These moduli transform
in the $(\mathbf{k},\mathbf{\overline N})$ representation of the $\mathrm{U}(k)\times \mathrm{U}(N)$ 
brane symmetry groups. By considering open strings with the opposite orientations we have other
moduli, $\bar w_{\dot \alpha}$, $\bar \mu_{\dot a}$, $\bar \mu_a$ and $\bar h_a$, all transforming in 
the $(\mathbf{\bar k},\mathbf{N})$ representation of $\mathrm{U}(k)\times \mathrm{U}(N)$.

\paragraph{Flavored sector:} In case we add $N_f$ D7 branes to realize $N_f$ fundamental flavors 
in the gauge theory, we also have flavored moduli. Due to the presence of eight directions with
mixed Neumann/Dirichlet boundary conditions, the physical spectrum of the D(--1)/D7 strings 
contains only one fermionic mode $\mu'$, which can be paired with an auxiliary field
$h$. These moduli transform in the $(\mathbf{k},\mathbf{\overline N_f})$ representation of
$\mathrm{U}(k)\times \mathrm{U}(N_f)$. The strings with opposite orientation give rise to
the moduli $\bar \mu'$ and $\bar h$ transforming in the 
$(\mathbf{\bar k},\mathbf{N_f})$ representation of $\mathrm{U}(k)\times \mathrm{U}(N_f)$.

\paragraph{} All these moduli and their properties are collected in Table~1 where we have organized them in three
sets. The first one is always present in $\mathcal N=2$ gauge theories, and thus we call it ``gauge''.
The second one has to be considered for $\mathcal N=4$ or $\mathcal N=2^*$ theories, 
that is when one has $\mathcal N=2$ massless or massive matter in the adjoint representation. 
For this reason in the first column of Table~1
we have used the terminology ``adjoint matter'' for this set of moduli. 
Finally, the third group is needed when there are D7 branes and thus matter in the fundamental representation.
As we can see from the second column of Table~1, all instanton moduli but $\chi$ can be arranged
into doublets $(\phi_s,\psi_s)$ of a $\epsilon$-deformed BRST charge such that
\begin{equation}
 Q \phi_s =\psi_s~, \qquad Q \psi_s= Q^2 \phi_s =\lambda_s \,\phi_s
\end{equation}
with $\lambda_s$ being the eigenvalues of $Q^2$ under action of the Cartan subgroup of the full symmetry group, namely the gauge group U($N$), the instanton symmetry U($k$), the flavor symmetry
U($N_f$), if present, and the residual SU(3)$^3$ Lorentz symmetry, chosen in such a way that invariant points in the moduli space are finite and isolated.
In the D brane realization of the gauge theory, the eigenvalues $\lambda_s$ are given by the distance 
in the $z$-plane between the two branes at the ends of the corresponding open string, 
shifted by the SU(2)$^3$ weights of the given mode. The positions of the instantons along 
the overall transverse $z$-plane are parametrized by the eigenvalues $\chi_I$ (with $I=1,..k$) 
of the scalar $\chi$. Similarly, we denote by $a_u$ (with $u=1,\ldots,N$) and $m_f$ (with 
$f=1,\ldots,N_f$) 
the positions of the D3 and D7 branes respectively along this plane. Using these notations
and the transformation properties of the various moduli, one can easily determine the
$Q^2$-eigenvalues $\lambda_s$ as reported in the last column of Table~1.
Note that complex conjugate fields have opposite eigenvalues and therefore, from now on, we can
concentrate on the holomorphic components, {\it i.e.} we take the plus signs in the
$\lambda_s$'s.
\begin{table}[h]
\begin{equation*}
\begin{array}{|c|c|c|c|c|l|}
\hline
& (\phi_s,\psi_s)  & (-1)^{F_s}  &\mathcal G & \mathrm{SU}(2)^3 &
 ~~~~~~~~~~~~~~\lambda_s \\
\hline \hline
{\mathrm{gauge}} &\phantom{\vdots}(B_\ell,M_\ell ) 
& +   & ~~({\bf k } {\mathbf{\bar k}},{\mathbf 1}, {\mathbf 1} ) ~~
& ~({\mathbf 2},{\mathbf 1},{\mathbf 2}) ~  
& ~\chi_{IJ}\pm \epsilon_{1}\,,\,\chi_{IJ}\pm \epsilon_{2}  \\
 & \phantom{\vdots}(\lambda_{c} ,D_c) & -   & ~~({\mathbf k } {\mathbf{\bar k}},{\mathbf 1}, 
{\mathbf 1} )~~  &
 ~({\mathbf 1},{\mathbf 1},{\mathbf 3})~   & ~  \chi_{IJ}\,,\, \chi_{IJ}\pm (\epsilon_1+\epsilon_2) \\
 & \phantom{\vdots}(\bar \chi ,\eta) & +   &~~ ({\mathbf k } {\mathbf{\bar k}},{\mathbf 1}, 
{\mathbf 1}) ~~ &
 ~({\mathbf 1},{\mathbf 1},{\mathbf 1})~   &  ~ \chi_{IJ}  \\
& \phantom{\vdots}
(w_{\dot \alpha},  \mu_{\dot \alpha}) & +  &~~({\mathbf k } ,{\mathbf{\overline N}}, {\mathbf 1} ) ~~  &  
~({\mathbf 1},{\mathbf 1},{\mathbf 2})~  & ~\chi_I-a_u\pm\ft12(\epsilon_1+\epsilon_2) \\
&\phantom{\vdots}(\bar w_{\dot \alpha},  \bar \mu_{\dot \alpha}) & +  &~~(\mathbf{\bar k } ,{\mathbf  N}, {\mathbf 1} )~~   
& ~ ({\mathbf 1},{\mathbf 1},{\mathbf 2})~  &~ a_u-\chi_I\pm\ft12(\epsilon_1+\epsilon_2) \\
 \hline\hline
{\rm adj.~ matter} & \phantom{\vdots}(B_{\dot \ell},M_{\dot \ell} ) & +  & ({\mathbf k } 
{\mathbf{\bar k}},{\mathbf 1}, {\mathbf 1} ) & ({\mathbf 1},{\mathbf 2},{\mathbf 2})   
& ~ \chi_{IJ}\pm \epsilon_{3}\,,\,\chi_{IJ}\pm (\epsilon_1+\epsilon_2+\epsilon_{3}) \\
& \phantom{\vdots}(\lambda_{m'} ,D_{m'}) & -   &  ({\mathbf k } {\mathbf {\bar k}},{\mathbf 1},
 {\mathbf 1} )  &    ({\mathbf 2},{\mathbf 2},{\mathbf 1})        &
~\chi_{IJ}\pm (\epsilon_1+\epsilon_3), \chi_{IJ}\pm (\epsilon_2+\epsilon_3) \\
& \phantom{\vdots}(\mu_{ a},  h_{  a}) & -  & ({\mathbf k } ,{\mathbf {\overline N}}, {\mathbf 1} ) &
({\mathbf 1},{\mathbf 2},{\mathbf 1})
 & ~\chi_I-a_u\pm \ft12(\epsilon_1+\epsilon_2+2\epsilon_3) \\
 & \phantom{\vdots}(\bar \mu_{ a},  \bar h_{ a}) & -  & ({\mathbf{\bar k} } ,{\mathbf  N}, {\mathbf 1} 
) &({\mathbf 1},{\mathbf 2},{\mathbf 1})
 & ~ a_u-\chi_I\pm \ft12(\epsilon_1+\epsilon_2+2\epsilon_3) \\
 \hline\hline
\phantom{\vdots}{\rm fund. ~matter} & (\phantom{\vdots}\mu',h) & -  &
( {\mathbf k }, {\mathbf 1}, {\mathbf {\overline N_f}}) & ({\mathbf 1},{\mathbf 1},{\mathbf 1})    & ~\chi_I-m_f   \\
& (\phantom{\vdots} \bar \mu', \bar h) & -  &( {\mathbf{ \bar k} }, {\mathbf 1}, {\mathbf{N_f}}) & 
({\mathbf 1},{\mathbf 1},{\mathbf 1})    & ~m_f-\chi_I \\
\hline
\end{array}
\end{equation*}
\label{modulitable}
\caption{Instanton moduli for ${\mathcal N}=2$ SU($N$) gauge theories. 
The various columns display the $Q$-multiplets, the spin-statistics $(-1)^{F_s}$ of the highest weight state $\phi_s$ in the multiplet ($F_s=0$ or 1 depending 
whether $\phi_s$ is bosonic or fermionic),
the transformation properties with respect to the brane symmetry group $\mathcal G=\mathrm{U}(k)\times
\mathrm{U}(N)\times\mathrm{U}(N_f)$ and the reduced Lorentz group $\mathrm{SU}(2)^3$, and finally
the $Q^2$-eigenvalues $\lambda_s$ (we have introduced the notation $\chi_{IJ}=\chi_I-\chi_J$). Note that the parameter $\epsilon_3$ plays the r\^ole of mass deformation for the adjoint matter multiplet.}
\end{table}

\subsection{Multi-instanton partition functions}

The instanton partition function is defined by the integral
\begin{equation}
 Z_{\mathrm{inst}}=\int d{\mathcal M_k} \,\ee^{-S_{\mathrm{inst}}}
=\sum_{k=0}^\infty \frac{q^k}{k!} 
\int \prod_{I=1}^k \frac{d\chi_I}{2\pi \ii}\,z_k
\label{zinst}
\end{equation}
with $z_0=1$ and
\begin{equation}
 z_k=z_k^{\mathrm{gauge}}\, z_k^{\mathrm{matter}} =\,(-1)^k \frac{\cV}{\det Q^2} 
= {\cV}\,(-1)^k \prod_s \lambda_s^{(-1)^{F_s+1}}
\label{det}
\end{equation}
where
\begin{equation}
\cV =   \prod_{IJ} \chi'_{IJ} 
\label{vandermonde}
\end{equation}
is the Vandermonde determinant resulting from the diagonalization of the $\chi$ matrix and $\chi'_{IJ}=\chi_{IJ}+\delta_{IJ}$. 
In (\ref{det}) the product index $s$ runs over the holomorphic field components. 

The partition function (\ref{zinst}) is computed by closing the integrals over the $\chi_I$'s
in the upper half plane with the following pole prescription \cite{Moore:1998et}
\begin{equation}
   1\gg {\mathrm{Im}}\epsilon_1 \gg {\mathrm{Im}}\epsilon_2 \gg \cdots >0~.
\label{pole}
\end{equation}
These integrals receive contributions from the poles of $z_k$, which are nothing but the critical points of $Q^2$. 
The poles of $z_k$ can be put in one-to-one correspondence with an $N$-array of Young tableaux
$Y=\{ Y_u \}$ (with $u=1,\ldots,N$) containing a total number of $k$ boxes 
and are given by
\begin{equation} \label{poles}
\chi^Y_{I}=
a_u+(i-\ft12) \epsilon_1+(j-\ft12) \epsilon_2
\end{equation}
with $(i,j)$ running over the rows and columns of the Young tableau $Y_u$. 
Exploiting this correspondence, the instanton partition function can then be written 
as a sum over Young tableaux arrays evaluated at the critical points, namely
\begin{equation}
Z_{\mathrm{inst}}= \sum_{Y}^\infty q^{|Y|}\, Z_Y~, \qquad  Z_Y=  
{\cV}(\chi_I^Y)\,\prod_s    \lambda_s(\chi_I^Y)^{(-1)^{F_s+1}} 
\end{equation}
with $|Y|=k$ being the total number of boxes in $Y$, and $\lambda_s(\chi_I^Y)$ and
${\cV}(\chi_I^Y)$ being respectively the $Q^2$-eigenvalues and Vandermonde determinant evaluated 
at the pole specified by the Young tableaux array $Y$.
It is convenient to introduce the notion of  character ${\bf T}$, given by the trace (rather than the determinant) over the instanton moduli space of $Q^2$
\begin{equation}
\mathbf{T}=\sum_{s}  (-1)^{F_s} \ee^{\ii \lambda_s}~.
\label{chardef}
\end{equation}
Even if the information in $\mathbf{T}$ and $Z_Y$ is completely equivalent, the character 
typically shows up cancellations between boson and fermionic contributions 
in $Z_Y$ in a more efficient way. 
 
The non-perturbative prepotential is identified with the free energy of the system
\begin{equation}
 F_{\mathrm{n.p.}}= -\lim_{\epsilon_1,\epsilon_2 \to 0}
\epsilon_1 \epsilon_2  \log  Z_{\mathrm{inst}}= \sum_{k=1}^\infty q^k F_k
\label{prepotinst1}
\end{equation}
while the chiral correlators of the gauge theory are given by~\cite{Fucito:2011pn}
\begin{equation}
 \begin{aligned}
  \big\langle \tr\,\ee^{z\,\Phi}\big\rangle = 
\big\langle \tr\,\ee^{z\,a}\big\rangle
-\frac{1}{Z_{\mathrm{inst}}}\,\sum_{k=1}^\infty
\frac{q^k}{k!} \int \prod_{I=1}^k \frac{d\chi_I}{2\pi \ii}\,z_k\,\cO(z,\chi_I)  
 \end{aligned}
\label{chiralcorr}
\end{equation}
with
\begin{equation}
 \label{Ochi}
\cO(z,\chi_I)= \sum_I  \ee^{z\chi_I}\big(1-\ee^{z\epsilon_1}\big)\big(1-\ee^{z\epsilon_2}
\big)~.
\end{equation}
{From} these equations it is not difficult to show that $\big\langle 
\tr\,{\Phi}^2\big\rangle$ is related to the prepotential
via the Matone's relation
\begin{equation}
\big\langle \tr\,{\Phi}^2\big\rangle
=2q\frac{dF}{dq}=
\sum_{u=1}^N a_u^2+2\sum_{k=1}^\infty k\,q^k F_k ~.
\end{equation}
Finally the matrix of gauge couplings is defined  as
\begin{equation}
2\pi\ii\,\tau^{uv}= \frac{\partial^2 F}{\partial a_u \partial a_v} ~.
\label{tauf}
\end{equation}
We now discuss a few specific examples.

\subsection{U($N$) with adjoint or fundamental matter}
We begin by considering the case of the U($N$) theory with adjoint matter, namely the so-called
$\mathcal N=2^*$. Collecting the $\lambda_s$ eigenvalues of the various fields from Table~1 
and taking into account the Vandermonde
determinant, we find
\begin{equation}
 \begin{aligned}
z^{\mathrm{gauge}}_k &= (-1)^{k}
\prod_{I,J}^k  \left[ 
\frac{ \chi'_{IJ} ( \chi_{IJ}+\epsilon_1+\epsilon_2)}{
(\chi_{IJ}+\epsilon_1)(\chi_{IJ}+\epsilon_2)}\right]
 \prod_{I=1}^k \prod_{u=1}^N\left[  \frac{1}{-(\chi_I-a_u)^2+\ft{(\epsilon_1+\epsilon_2)^2}{4}}  
 \right]~, \\
z^{\mathrm{adj.matt.}}_k &=
\prod_{I,J}^k \left[  \frac{( \chi_{IJ}+\epsilon_1+\epsilon_3)( \chi_{IJ}+\epsilon_2+\epsilon_3)}{
(\chi_{IJ}+\epsilon_3)(\chi_{IJ}+\epsilon_1+\epsilon_2+\epsilon_3)}\right]
\prod_{I=1}^k \prod_{u=1}^N  \left[ -(\chi_I-a_u)^2
   +\ft{(\epsilon_1+\epsilon_2+2\epsilon_3)^2}{4} \right]~.
\end{aligned}
 \label{zk}
\end{equation}
As we explained before, the evaluation of the residues of  (\ref{zk}) can be efficiently performed by first computing 
the character ${\bf T}$ defined in (\ref{chardef}).
 Collecting the eigenvalues of the various moduli in table \ref{modulitable} for ${\cal N}=2^*$ one finds 
\begin{equation}
 \mathbf{T}_{\cN=2^*}= \mathbf{T}_{\mathrm{gauge}}+\mathbf{T}_{\mathrm{adj. matt.}}
\end{equation}
with
\begin{equation}
 \begin{aligned}
  &\mathbf{T}_{\mathrm{gauge}}=-V\, V^* \, (1-T_1)\,(1-T_2)+V^* \, W +V \, W^* \, T_1\, T_2~,\\
&\mathbf{T}_{\mathrm{adj.matt.}}=-T_3\,\mathbf{T}_{\mathrm{gauge}}
 \end{aligned}
\label{chiy}
\end{equation}
where
\begin{equation}
 V=\sum_{I=1}^k \ee^{\ii \left(\chi_I-\ft{\epsilon_1+\epsilon_2}{2}\right) }~,\qquad
W= \sum_{u=1}^N T_{a_u}
\label{vecspaces}
\end{equation}
and $V^*$ and $W^*$ are their complex conjugates.
In these expressions we have used the following notation
\begin{equation}
 T_1=\ee^{\ii\epsilon_1}~,~~~T_2=\ee^{\ii\epsilon_2}~,~~~T_3=\ee^{\ii\epsilon_3}~,~~~
T_{a_u}=\ee^{\ii a_u}~.
\label{ts}
\end{equation}
\eq{chiy} can be recovered from the exact sequence associated with an instanton (see Ref.~\cite{donaldson} for a complete treatment and Ref.s~\cite{Flume:2004rp,Bruzzo:2002xf,Shadchin:2004yx} for an exposition connected to this application). If we use the definitions (\ref{vecspaces}) and (\ref{ts}) in (\ref{chiy}) and take the
product of the exponents, we can recover (\ref{zk}). On the other hand, we can 
substitute the values of the poles of this latter equation in the definition of $V$ and get from (\ref{poles})
\begin{equation}
V(Y)=\sum_{u=1}^N \sum_{i,j\in Y_u}T_{a_u}T_1^{i-1}T_2^{j-1}~.
\end{equation}
Thus $V$ can be written as a polynomial in the $T$'s which
encodes the D(--1) positions labeling the Young tableaux array 
$Y=\{ Y_u \}$. The various terms in $V$ and $V^*$ are associated to open string 
modes starting or ending on the D-instantons. Similarly $W$ and $W^*$ account for
open strings starting or ending on D3 branes. For instance the
four terms $V\, V^* \,(T_1+T_2+T_3+T_1 T_2 T_3)$ come from the neutral moduli
$B_\ell$ and $B_{\dot\ell}$, 
while the terms $V^* \, W$ and $V \, W^*\,T_1 \,T_2$ come from the charged moduli $w_{\dot\alpha}$
and $\bar w_{\dot\alpha}$. The remaining terms come with negative signs and originate from the
auxiliary fields. They realize a sort of generalized ADHM constraints~\footnote{
We assign $F=0$ to fields entering in the character with positive sign and $F=1$ otherwise.}.

{From} these expressions that are valid in the $\cN=2^*$ theory, one obtains 
the pure $\cN =2$ theory by decoupling the adjoint matter sending its mass to infinity. 
This corresponds to the limit $T_3 \to 0$. 
On the other hand, we can add fundamental matter by introducing $N_f$ D7 branes 
whose contribution to the character is given by
\begin{equation}
 \mathbf{T}_{\mathrm{fund.matt.}} = -\sum_{f=1}^{N_f} V \, T_{m_f}
\end{equation}
with $T_{m_f}=\ee^{-\ii \,\left( m_f-\frac{\epsilon_1+\epsilon_2}{2}\right)}$, $m_f$ being the masses of the fundamental flavors which coincide with the D7 brane positions. 
This leads to the contribution
\begin{equation}
z_k^{\mathrm{fund. matt.}} =\prod_{I=1}^k \prod_{f=1}^{N_f} (\chi_I-m_f)
\end{equation}
in the instanton partition function.

\paragraph{$\bullet$ Example: U(1) plus adjoint matter.}
As a first example, let us consider a U(1) gauge theory with adjoint matter. To emphasize the r\^ole
of $\epsilon_3$ as mass parameter for the adjoint multiplet, we rename it $m$ and use the 
symbol $T_m$ in place of $T_3$. We then find for the first few tableaux
\begin{equation}
\begin{aligned}
{\mathbf T}(\Yfund)&= (1-T_m)(T_{1}+T_{2})~,\\
{\mathbf T}(\Ysymm)&= (1-T_m)\left(T_{1} +T_{2}+T_{1}^2+\frac{T_2}{T_1}\right)
\end{aligned}
\label{tts}
\end{equation}
with a similar contribution for ${\mathbf T}(\Yasymm)$ obtained by exchanging
$\epsilon_1 \leftrightarrow \epsilon_2$ in the second line. Note that all $m$-independent
terms arising from the gauge character come with strictly positive coefficients. The opposite is 
true for the terms coming from the matter fields proportional to $T_m$. 
Collecting the eigenvalues from (\ref{tts}) one obtains the following contributions 
to the partition function
\begin{equation}
\begin{aligned}
Z_{\Yfund} &=\frac{( m+\epsilon_1)( m+\epsilon_2)}{\epsilon_1 \epsilon_2}~,\\
Z_{\Ysymm}&= \frac{( m+\epsilon_1)( m+\epsilon_2)( m+\epsilon_2-\epsilon_1)
( m+2\epsilon_1)}{2 \epsilon_1^2 \epsilon_2( \epsilon_2-\epsilon_1)}~,
 \end{aligned}
\label{zu1}
\end{equation}
which lead to the non-perturbative prepotential
\begin{equation}
F_{\mathrm{n.p.}}= -\lim_{\epsilon_1,\epsilon_2 \to 0}
\epsilon_1 \epsilon_2 \Big[ q \, Z_{\Yfund} +
q^2 \Big(Z_{\Ysymm} + Z_{\Yasymm}-\frac12 Z_{\Yfund}^2\Big)\Big]
= -q \,m^2-\frac{3}{2} \,q^2 \,m^2+\cdots~.
\end{equation}

\paragraph{$\bullet$ Example: Pure SU(2) gauge theory.}
For the pure SU(2) gauge theory we take $T_{a_1}=T_{a_2}^{-1}=T_a$.
The contributions of the first few tableaux are
\begin{equation}
 \begin{aligned}
  {\mathbf T}(\Yfund,\bullet) &=T_{1}+T_{2}+T_{-2a}+T_1 T_2 T_{2a} ~,\\
{\mathbf T}(\Ysymm,\bullet) &=T_{1}  +T_{1}^2+T_{2}+\frac{T_{2}}{T_1}+T_{-2a}+T_1 T_2 T_{2a}
  +T_1^2 T_2 T_{2a}+ \frac{T_{-2a}}{T_1}~,\\
{\mathbf T}(\Yfund,\Yfund) &=2T_{1}  +2T_{2}+T_{1} T_{2a}+T_{2} T_{2a}+
    T_{1} T_{-2a}+T_{2} T_{-2a}~.
 \end{aligned}
\end{equation}
Similar expressions for ${\mathbf T}(\bullet,\Yfund)$ and ${\mathbf T}(\bullet,\Ysymm)$ are
obtained from the first two lines by replacing $a\to -a$, while ${\mathbf T}(\Yasymm,\bullet)$
is obtained from the second line by exchanging $\epsilon_1 \leftrightarrow \epsilon_2$ and 
${\mathbf T}(\bullet,\Yasymm)$ is obtained from the second line after replacing $a\to -a$ and
$\epsilon_1 \leftrightarrow \epsilon_2$ simultaneously. These results lead to the instanton
partition functions
\begin{equation}
 \begin{aligned}
  Z_{(\Yfund,\bullet)}&=- \frac{1}{2a\,\epsilon_1 \epsilon_2 (2a+\epsilon_1+\epsilon_2)}~,\\
Z_{(\Ysymm,\bullet)}&= \frac{1}{4a \epsilon_1^2 \epsilon_2(\epsilon_2-\epsilon_1)
 (2a+\epsilon_1+\epsilon_2)(2a+2\epsilon_1+\epsilon_2)(2a+\epsilon_1)}~,\\
Z_{(\Yfund,\Yfund)}&=\frac{1}{\epsilon_1^2 \epsilon_2^2 (4a^2-\epsilon_1^2)
 (4a^2-\epsilon_2^2)}
 \end{aligned}
\end{equation}
from which we find the prepotential
\begin{equation}
  F_{\mathrm{n.p.}}
    = \frac{q}{2 a^2} +\frac{5q^2}{64a^6}+\cdots~.
\end{equation}

\subsection{U($N_0$)$\times$U($N_1$) quiver gauge theory}

As explained in the main text, we realize this quiver gauge theory by placing $N_0$ fractional D3-branes of type 0 and $N_1$ fractional D3 branes of type 1 at the $\C^2/\Z_2$ singularity.
The moduli space of the U($N_0$) group is parametrized by the vacuum expectation values 
$a_u$ ($u=1,\ldots,N_0$), while that of the U($N_1$) factor by $b_v$ ($v=1,\ldots,N_1$).
There are also two types of fractional instantons whose numbers we denote by $k_0$ and $k_1$
and their positions in the transverse space by $\chi_{0I}$  and $\chi_{1J}$ with $I=1,\ldots,k_0$ and $J=1,\ldots,k_1$.

The $\Z_2$ orbifold induces the decompositions
\begin{equation}
 V=V_{0}+V_{1}~,~~~ W =W_{0}+W_{1}~,
\end{equation}
with
\begin{equation}
 \begin{aligned}
  V_0 &=  \sum_{I=1}^{k_0} \ee^{\ii \left(\chi_{0I}-\ft{\epsilon_1+\epsilon_2}{2}\right) }~,\qquad
W_0=\sum_{u=1}^{N_0} T_{a_u}~,\\ 
V_1 &= \sum_{J=1}^{k_1} \ee^{\ii \left(\chi_{1J}-\ft{\epsilon_1+\epsilon_2}{2}\right) }~,\qquad
W_1=\sum_{v=1}^{N_1} T_{b_u}~,
 \end{aligned}
\label{quiverspace}
\end{equation}
on which the $\Z_2$ action is (see Ref.~\cite{Fucito:2004gi} for a detailed treatment of this case)
\begin{equation}
 V_1 \to -V_1~,\qquad W_1\to -W_1~, \qquad T_3\to -T_3~.
\label{z2act}
\end{equation}
The character associated to the quiver theory follows then from that of the U($N_0+N_1$)
theory with adjoint matter by keeping only the terms invariant under (\ref{z2act}). 
This prescription leads to
 \begin{equation}
\mathbf{T}_{\mathrm{quiver}}=\sum_{a=0,1}  \left(\mathbf{T}_{\mathrm{gauge},a}+\mathbf{T}_{\mathrm{bifund. matt.},a}\right)
  \label{quivercharacter1}
 \end{equation}
where
\begin{equation}
\begin{aligned}
\mathbf{T}_{\mathrm{gauge},a} &=-V_a\, V_a^* \, 
(1-T_1)\,(1-T_2)+W_a\, V_a^*  +V_a \, W_a^* \, T_1\, T_2~,\\
\mathbf{T}_{\mathrm{bifund. matt.},a}&= T_3\,  \Big[
V_a\, V_{a+1}^* \, (1-T_1)\,(1-T_2)-W_a\, V_{a+1}^*  -V_a \, W_{a+1}^* \, T_1\, T_2\Big]
   \end{aligned}
\label{quivercharacter}
  \end{equation}
with subscripts understood modulo 2. 
Similarly, the multi-instanton partition function for the quiver theory is given by the product of the eigenvalues in (\ref{quivercharacter}) that are nothing but
the $\Z_2$ invariant components in the ${\cal N}=2^*$ theory. More precisely,
the terms surviving the projection are associated either to strings connecting two branes 
of the same type (gauge components) or to strings connecting branes of opposite types 
which give rise to bi-fundamental matter.
Proceeding as described in the previous subsections, we get
\begin{equation}
Z_{\mathrm{inst}}=
\sum_{k_0,k_1=0}^\infty \, \frac{q_0^{k_0}}{k_0!} \,\frac{q_1^{k_1}}{k_1!}\, 
\int \prod_{I=1}^{k_0} \frac{d\chi_{0I}}{2\pi \ii}
\prod_{J=1}^{k_1} \frac{d\chi_{1J}}{2\pi \ii}~ 
z^{\mathrm{quiver}}_{k_0,k_1} 
\label{zkquiver}
\end{equation}
with
\begin{equation}
 z^{\mathrm{quiver}}_{k_0,k_1} =z^{\mathrm{gauge}}_{k_0}
~z^{\mathrm{gauge}}_{k_1}
~z^{\mathrm{bifund.matt}}_{k_0,k_1}~
\end{equation}
Here $z^{\mathrm{gauge}}_{k_0}$ and $z^{\mathrm{gauge}}_{k_1}$ are as in (\ref{zk}) and depend
on the parameters of the two gauge groups, while $z^{\mathrm{bifund.matt}}_{k_0,k_1}$ is obtained
from $z^{\mathrm{adj.matt}}_{k}$ given in (\ref{zk}) with $k=k_0+k_1$ and $N=N_0+N_1$.
More explicitly we have
 \begin{equation}
  \begin{aligned}
   z^{\mathrm{bifund.matt}}_{k_0,k_1}
&=\prod_{I=1}^{k_0}\prod_{J=1}^{k_1} 
\frac{\left[(\chi_{0I}-\chi_{1J})^2-(\epsilon_1+\epsilon_3)^2\right]
\left[(\chi_{0I}-\chi_{1J})^2-(\epsilon_2+\epsilon_3)^2\right] }
{\left[(\chi_{0I}-\chi_{1J})^2-\epsilon_3^2\right]
\left[(\chi_{0I}-\chi_{1J})^2-(\epsilon_1+\epsilon_2+\epsilon_3)^2\right]}\\
&~~~~\times \prod_{I=1}^{k_0}\prod_{v=1}^{N_1}\left[ -(\chi_{0I}-b_v)^2 +\frac{(\epsilon_1+\epsilon_2+2\epsilon_3)^2}{4}\right]\\
&~~~~\times \prod_{J=1}^{k_1}\prod_{u=1}^{N_0}\left[ -(\chi_{1J}-a_u)^2 +\frac{(\epsilon_1+\epsilon_2+2\epsilon_3)^2}{4}\right]
  \end{aligned}
 \end{equation}
with $z^{\mathrm{bifund.matt}}_{0,0}=1$. 
The poles of the integrand of the instanton partition function (\ref{zkquiver}) can be explicitly
determined using the previous formulae and, as before, turn out to be in one-to-one correspondence 
with Young tableaux arrays. Substituting them in (\ref{quiverspace}), one gets 
\begin{equation}
V_0(Y) =\sum_{u=1}^{N_0}\,\sum_{i,j \in Y_u} T_{a_u}\,T_1^{i-1}\,T_2^{j-1}~,~~~~ V_1(Y)=\sum_{v=1}^{N_1}\, \sum_{i,j\in Y_v} T_{b_v}\,T_1^{i-1}\,T_2^{j-1}
\end{equation}
which, once inserted in (\ref{quivercharacter1}), allow to perform computations 
similar to those described in the previous paragraphs.

The chiral correlators of the quiver gauge theory are given by insertions of $\chi_{0I}$ or 
$\chi_{1J}$ inside the instanton partition function. More precisely the correlators of the
U($N_0$) node are given by
 \begin{equation}
 \big\langle \tr_{N_0} \ee^{z \Phi}\big\rangle = \big\langle \tr_{N_0} \ee^{z a} \big\rangle  
- \frac{1}{Z_{\mathrm{inst}}}
\sum_{k_0,k_1=0}^\infty \, \frac{q_0^{k_0}}{k_0!} \,\frac{q_1^{k_1}}{k_1!}\, 
\int \prod_{I=1}^{k_0} \frac{d\chi_{0I}}{2\pi \ii}
\prod_{J=1}^{k_1} \frac{d\chi_{1J}}{2\pi \ii}~ 
z^{\mathrm{quiver}}_{k_0,k_1}~{\cO}(z,\chi_{0I})
\label{trephi0}
 \end{equation}
where $\cO(z,\chi_{0I})$ is defined in (\ref{Ochi}). A similar formula holds for $\big\langle 
\tr_{N_1} \ee^{z \Phi}\big\rangle$ with the replacement ${\cO}(z,\chi_{0I})\leftrightarrow
{\cO}(z,\chi_{1I})$.
Again the prepotential is related to $\big\langle \tr_{N_j}{\Phi}^2\big\rangle$ via a
generalized Matone relation, namely
\begin{equation}
\big\langle \tr_{N_j} {\Phi}^2\big\rangle
= 2 q_j\frac{d{F}}{dq_j}~~\quad\mbox{for}~j=0,1~.
\end{equation}
We finally remark that any chiral correlator in the quiver theory receives contributions from
both types of instantons and therefore is a double series expansion in $q_0$ and $q_1$.

\paragraph{$\bullet$ Example: SU(2)$\times$SU(2).}
We now consider the case $N_0=N_1=2$. Here we take $a_1=-a_2=a$, $b_1=-b_2=b$ and 
$\epsilon_3=0$. Moreover we set $\epsilon_2=-\epsilon_1$. The contributions to the character 
for the first few Young tableaux are
\begin{equation}
 \begin{aligned}
 \phantom{\Big|}&\mathbf{T}_{\mathrm{quiver}}(\Yfund,\bullet\big|\bullet,\bullet) =
T_1 +T_{2a}-T_{a+b}-T_{a-b}+\mathrm{h.c.}~,\\
\phantom{\Big|}&\mathbf{T}_{\mathrm{quiver}}(\bullet,\bullet\big|\Yfund,\bullet)=T_1 +T_{2b}-T_{a+b}-T_{a-b}
+\mathrm{h.c.}~,\\
\phantom{\Big|}&\mathbf{T}_{\mathrm{quiver}}(\Yfund,\bullet\big|\Yfund,\bullet) =
2\,T_1+T_{2a} +T_{2b}-2 T_{a+b}-T_{a-b+\epsilon_1}-T_{a-b-\epsilon_1} +\mathrm{h.c.}~,\\
\phantom{\Big|} &\mathbf{T}_{\mathrm{quiver}}(\Yfund,\bullet\big|\bullet,\Yfund) =
2\,T_1+T_{2a} +T_{2b}
-2 T_{a-b}-T_{a+b+\epsilon_1}-T_{a+b-\epsilon_1} +\mathrm{h.c.}~.
 \end{aligned}
\label{tsy}
\end{equation}
The remaining four Young tableaux array obtained from those above by exchanging the order
of boxes and bullets at the two sides of the bracket simultaneously give identical results. Summing up all contributions, 
one finds (up to order $k_0,k_1=1$)
\begin{equation}
 \begin{aligned}
Z_{\mathrm{inst}} &= 1+ \frac{(a^2-b^2)^2}{2\epsilon_1^2}
\Big[\frac{q_0}{a^2} + \frac{q_1}{b^2}\Big] \\
&~~+\frac{q_0 q_1}{8 \epsilon_1^4 a^2 b^2}\,\Big\{
(a+b)^4 \big[(a-b)^2-\epsilon_1^2\big]^2+ (a-b)^4 \big[(a+b)^2-\epsilon_1^2\big]^2\Big\}+\cdots
 \end{aligned}
\end{equation}
which leads to the non-perturbative prepotential \cite{Fucito:2004gi}
\begin{equation}
 F_{\mathrm{n.p.}} = (a^2-b^2)^2\left(\frac{q_0}{2a^2}+\frac{q_1}{2b^2} \right)
 -q_0\,q_1\,\frac{(a^2+b^2)(a^2-b^2)^2}{2a^2\,b^2} +\cdots~.
\end{equation}
Now let us consider the chiral correlators. {From} (\ref{trephi0}), it is possible to
show that
 \begin{equation}
 \big\langle \tr_{2} \log (z-\phi) \big\rangle 
= \log (z^2-a^2) + \lim_{\epsilon_1\to0}\frac{1}{Z_{\mathrm{inst}}} 
\sum_{Y} q_0^{k_0}q_1^{k_1}\, Z_Y \,\tr_{k_0}
\log \Big( 1-\frac{\epsilon_1^2}{(z-\chi_0^Y)^2}\Big)~.
  \end{equation}
Working out the first few instanton contributions, we obtain
\begin{equation}
 \begin{aligned}
 \big\langle \tr_{2} \log (z-\phi) \big\rangle  &=\log (z^2-a^2)
-q_0\,\frac{(a^2-b^2)^2\,(a^2+z^2)}{2a^2\,(a^2-z^2)^2}\\
&~~\qquad\qquad+q_0\,q_1\,\frac{(a^2-b^2)^2\,(a^2+b^2)\,(a^2+z^2)}
{2a^2b^2 \,(a^2-z^2)^2}+\cdots~,
 \end{aligned}
\end{equation}
and expanding for large $z$, we get
\begin{eqnarray}
\big\langle \tr_{2} \log (z-\phi) \big\rangle  &=&\log z^2 -\sum_{\ell=1}\frac{\big\langle
\tr_2\phi^\ell\big\rangle}{\ell\,z^\ell}\\
&=&\log z^2
-\frac{1}{2z^2}\,\Big[2a^2+q_0\,\frac{(a^2-b^2)^2}{a^2}-
q_0\,q_1\,\frac{(a^2-b^2)^2(a^2+b^2)}{a^2b^2}+\cdots\Big]\nonumber\\
&&~~-\frac{1}{4z^4}\,\Big[2a^4+6q_0\,(a^2-b^2)^2-
q_0\,q_1\,\frac{6(a^2-b^2)^2\,(a^2+b^2)}{b^2}+\cdots\Big]+\cdots
\nonumber
\end{eqnarray}
from which we can read the expressions of the chiral ring elements of the first SU(2) factor.
Similar expressions can be found for the correlators of the second SU(2) by simply replacing $a\leftrightarrow b$ and $q_0\leftrightarrow q_1$.

\subsection{SU($N$) with $2N$ hypermultiplets in the special vacuum}
\label{sunspecial}
In Section~\ref{secn:gaugegrav} we have mentioned that 
the couplings (\ref{taq0}), (\ref{taq03}) and (\ref{taq0m4}) are compatible with
the same expressions obtained from the SW curve (\ref{curve}) for the SU($N$) theory
with $2N$ flavors in the special vacuum. In this subsection we give some details on how to compute such couplings.

To this aim let us first introduce the matrices
 \begin{equation}
  A_{i,{\ell}}=\frac{\partial a_i}{\partial u_{\ell+1}} 
=\frac{1}{2 \pi \ii} \oint_{\gamma_i}\omega_{\ell}(z)~,\qquad
B_{i,{\ell}} = \frac{\partial a_{Di}}{\partial u_{\ell+1}}
 =\frac{1}{2 \pi \ii} \oint_{\widetilde \gamma_i}\omega_{\ell}(z)
\label{periodsab} 
 \end{equation}
where $i,\ell=1,\ldots,N-1$, $u_\ell$ are the invariants defined in (\ref{PQ}) and
$ \omega_{\ell} =  - z^{N-1-\ell}\,\frac{dz}{y}$
are meromorphic differentials whose form can be found from (\ref{aad}), using (\ref{SWdifferentialnoi}) together with $\frac{\partial  P}{\partial u_{\ell+1}}=z^{N-1-\ell}$.
The integrals in (\ref{periodsab}) are calculated over a basis of cycles 
$(\gamma_i,\widetilde\gamma_j)$ on the $N$-cut complex plane
such that $\gamma_i \circ \widetilde\gamma_j=\delta_{ij}$. 

The gauge coupling matrix $\tau^{ij}$ then follows directly from (\ref{periodsab}); indeed
\begin{equation}
 \tau^{ij}=\frac{\partial {a_D}_i}{\partial a_j}=(B A^{-1} )^{ij}~.
\label{tauAB}
\end{equation}
In the special vacuum 
the $2N$ branching points are given by
\begin{equation}
 z_{2u-1}=\alpha_1 \,\omega^{u-1}~,\qquad
z_{2u}=\alpha_2 \,\omega^{u-1}
\end{equation}
with $u=1,\ldots,N$ and
 \begin{equation}
  \alpha_1 =\left(\frac{-u_N+g\, m^N}{1+g}\right)^{\frac{1}{N}}~,\qquad
\alpha_2 =\left(\frac{-u_N-g\, m^N}{1-g}\right)^{\frac{1}{N}}
 \end{equation}
The period integrals are computed as line integrals with the identifications 
\begin{equation}
\oint_{\gamma_i} \omega_{\ell}= 2\int_{z_{2i-1}}^{z_{2i}} \omega_{\ell} ~, \qquad   \oint_{\widetilde{\gamma}_i} \omega_{\ell}=2\sum_{j=i}^{N-1} 
\int_{z_{2j}}^{z_{2j+1}} \omega_{\ell}
\label{periods}
\end{equation}   
where the cycles $\gamma_i$ are around the cuts $[z_{2i-1},z_{2i}]$ and the dual cycles are defined 
by the condition $\gamma_i \circ \widetilde\gamma_j=\delta_{ij}$ (see Fig.~\ref{fig:cycles} for an
example of these cycles in the case $N=3$). With these positions, the integrals in (\ref{periodsab}) 
are now of the type
\begin{equation}
 I_{ab}^{(k)}=\int_{z_a}^{z_b}  \frac{x^k\,dx}{(x^N-\alpha_1^N)^{\frac12} \,
(x^N-\alpha_2^N)^{\frac12}}=I^{(k)}(z_b)-
 I^{(k)}(z_a)
\label{integral}
\end{equation} 
with $a,b=1,..2N$. The quantities in the right hand side are indefinite integrals given 
in terms of the Euler $\Gamma$-function and the hypergeometric function ${}_2 F_1$ according to
 \begin{eqnarray}
 I^{(k)}(x) &=&\int^x  \frac{z^k\, dz}{(z^N-\alpha_1^N)^{\frac12}  \,
(z^N-\alpha_2^N) ^{\frac12}} \\
 &=&\frac{  \sqrt{\pi} \,  x^{k+1}}{(k+1)\,(\alpha_1\alpha_2)^{\frac{N}{2}}} 
\frac{ \Gamma\left(\ft{N+k+1}{N}\right)}{ \Gamma \left(\ft{N+2(k+1)}{2N}\right) } \,  {}_2 F_1 \left(\ft{k+1}{N}, \ft12,\ft{N+2(k+1)}{2N}, \big(\ft{x^2 }{\alpha_1 \alpha_2}\big)^N\right)\nn
\end{eqnarray}
for $x^N\in \{ \alpha_{1}, \alpha_{2} \}$. 
We have checked these results with the microscopic instanton calculus described in the previous 
subsections and found complete agreement.

\section{String diagrams}
\label{secn:appa}
In this appendix we give some details on the disk diagrams that are relevant
for computing the instanton moduli action in our $\mathbb C^2/\mathbb Z_2$ orbifold model.

\subsection{Interaction among $t$ and the $\theta$ moduli}

The diagonal parts of the $B_{\ell}$
and $M^{\alpha\dot{a}}$ moduli, introduced in the previous appendix, represent the bosonic and fermionic Goldstone modes
of the supertranslations of the D3 brane worldvolume broken by the D-instantons,
and thus they can be identified with the $\cN=2$ chiral superspace coordinates.
For example, working in units of the string length $\ell_s=\sqrt{\alpha'}$, the
fermionic coordinates are $\theta^{\alpha\dot{a}}\sim
\tr\,M^{\alpha\dot{a}-}$.

We now derive the interaction of the twisted scalar $t$ with the $\theta$ moduli.
To this aim we first consider a disk diagram having D$(-1)$ boundary conditions (say of type 0)
with the insertion of a $b$ vertex in the interior and four $\theta$ vertices on the boundary:
\begin{equation}
\label{diagtheta}
A_b=\big\langle V_{\theta}V_{\theta}V_{\theta}V_{\theta}\,V_b\,\big\rangle~.
\end{equation}
To saturate the superghost disk anomaly, we put the $b$ vertex in the $(-1,-1)$ picture, two
$\theta$ vertices in the $-\frac{1}{2}$ picture and two vertices in the $+\frac{1}{2}$ picture.
These vertex operators are:
\begin{equation}
\begin{aligned}
&V_{b}^{(-1,-1)}=
\pi\,b\,\epsilon^{\dot{a}\dot{b}}\,S_{\dot{a}}(w)\tilde{S}_{\dot{b}}(\bar{w})\,
\Delta(w)\tilde{\Delta}(\bar{w})\,\ee^{\ii [p \bar Z(w,\bar{w})+ \bar p Z(w,\bar{w})]}\,\ee^{-\varphi(w)}\ee^{-\tilde{\varphi}(\bar{w})}~,\\
&V^{(-\frac{1}{2})}_\theta =\frac{\sqrt{2}}{2\pi}\,\theta^{\alpha\dot{a}}\,
S_{\alpha\dot{a}+}(x)\,\ee^{-\frac{1}{2}\varphi(x)}~,\\
&V^{(+\frac{1}{2})}_\theta=\frac{\sqrt{2}}{2\pi}\,
\theta^{\alpha\dot{a}}\,\partial\bar{Z}(x)\,S_{\alpha\dot{a}-}(x)\,\ee^{\frac{1}{2}\varphi(x)}~.
\end{aligned}
\label{verticesbs}
\end{equation}
In these definitions $\Delta$ and $\tilde{\Delta}$ are the twist and anti-twist fields
in the orbifolded directions, $S_{\dot{a}}$ is the spin field in these directions,
$S_{\alpha\dot{a}+}$ is the spin field in the entire ten-dimensional space and $Z(x)$
is the bosonic string field along the complex plane transverse to the branes.
The momentum $p$ in the $b$ vertex is taken to be non-vanishing only along
the transverse space.

Since the amplitude (\ref{diagtheta}) involves four $\theta$'s, in order to get a non-vanishing
result their spinor indices must be all different; thus we can choose a fixed value for them and then sum over cyclically inequivalent orderings. We can also make a definite choice for the
spinor indices of spin fields in the $b$ vertex since, after summing over cyclically
inequivalent orderings, the other possible choice yields the same result.
Therefore the amplitude (\ref{diagtheta}) becomes
\begin{equation}\label{amptheta}
A_b=k_0\,\frac{\pi}{g_s}\,\frac{b\,\theta^4}{4\pi^3}\,
\int\frac{d^2w\prod_{i=1}^4dx_i}{dV_{\mathrm{CKG}}}\,C_{\mathrm{tot}}
\end{equation}
where the factor of $k_0$ comes from the trace over the Chan-Paton indices of the D-instantons, the factor of $\frac{\pi}{g_s}$ accounts for the topological normalization of the
disk with D(--1) boundary conditions, and $dV_{\mathrm{CKG}}$ stands for the volume of the conformal Killing group. Finally, $C_{\text{tot}}$ is twice the correlator of the vertex operators summed over cyclically inequivalent orderings of the four open string vertices, namely
\begin{equation}\label{ctot}
C_{\text{tot}}=2(C_{1234}+C_{1243}+C_{1324}+C_{1342}+C_{1423}+C_{1432})~.
\end{equation}
Let us choose the spinor indices of the spin fields as reported in the following table:
\begin{center}
\begin{tabular}{|c|c|c|}
\hline
$S_{\alpha_1\dot{a}_1+}$&$+++-$&$+$\\
\hline
$S_{\alpha_2\dot{a}_2+}$&$---+$&$+$\\
\hline
$S_{\alpha_3\dot{a}_3-}$&$++-+$&$-$\\
\hline
$S_{\alpha_4\dot{a}_4-}$&$--+-$&$-$\\
\hline
$S_{\dot{a}}$&$~~~~~~+-$&$ $ \\
\hline
$\tilde {S}_{\dot{b}}$&$~~~~~~-+$&$ $ \\
\hline
\end{tabular}
\end{center}
Then, using standard CFT techniques (see for instance Ref.~\cite{Billo:2011uc} for further details)
we find
\begin{equation}
C_{1234}=-\ii\bar{p}^2\big[(x_1-x_2)(x_3-x_4)(x_1-\bar{w})(x_2-w)(x_3-w)(x_4-\bar{w})\big]^{-1}~.
\end{equation}
The correlators corresponding to other orderings can be similarly evaluated by placing the vertices
with given indices at permuted locations $x_i$. Summing over all the inequivalent
orderings, we then find
 \begin{equation}
A_b=-k_0\,\frac{\pi}{g_s}\,\frac{b\,\theta^4}{4\pi^3}\,3\ii \bar{p}^2 \int\frac{d^2w\prod_{i=1}^4dx_i}{dV_{\mathrm{CKG}}}\,
\,\frac{(w-\bar{w})^2}{\prod_{i=1}^4\lvert x_i-w\rvert^2}
= -k_0\pi\,\bar p^2\,\theta^4\,\frac{b}{g_s}~.
\label{ab}
\end{equation}

Let's now consider the disk amplitude describing the interaction of the R-R twisted scalar
$c$ and four $\theta$'s
\begin{equation}
A_c=\big\langle V_{\theta}V_{\theta}V_{\theta}V_{\theta}\,V_c\,\big\rangle~.
\label{diagtheta2}
\end{equation}
We take the R-R vertex in the $(-\frac{1}{2},-\frac{1}{2})$ picture, namely
\begin{equation}
V_{c}^{(-\frac{1}{2},-\frac{1}{2})}=
\pi\,g_s
\,(\ii\bar{p} c)\,\epsilon^{\alpha\beta}\,S_{\alpha-}(w)\tilde{S}_{\beta-}(\bar{w})
\Delta(w)\tilde{\Delta}(\bar{w})\,\ee^{\ii[p \bar Z(w,\bar{w})+ \bar p Z(w,\bar{w})]}
\,\ee^{-\frac{1}{2}\varphi(w)}\ee^{-\frac{1}{2}\tilde{\varphi}(\bar{w})}~.
\label{vc}
\end{equation}
Thus, to saturate the superghost anomaly we must take
three $\theta$ vertices in the $-\frac12$ picture and one in the
$+\frac12$ picture.
The calculation of the amplitude (\ref{diagtheta2}) proceeds very similarly to the previous case:
again we can fix the spinor indices and then sum over inequivalent orderings of the open string vertices, obtaining the following result
\begin{equation}
A_c=-\ii\pi\,k_0\,\bar p^2\,\theta^4\,c~.
\label{ac}
\end{equation}
Adding Eq.s~(\ref{ab}) and (\ref{ac}) and recalling that the contribution to the euclidean action is minus the amplitude, we find the following interaction term in the effective action
\begin{equation}
\ii \pi\,k_0\,\bar{p}^2\,\theta^4\,\big(c-\ii\frac{b}{g_s}\big)=
\ii \pi\,k_0\,\bar{p}^2\,\theta^4\bar{t}
\end{equation}
which is part of $-\ii\pi k_0 T$, as reported in the main text.

\subsection{Interaction among $t$ and many $\chi$'s}

Another interaction which plays a crucial r\^ole in our analysis is the one among the twisted scalar $t$ and many $\chi$ moduli. To compute this coupling, let us first consider an instanton disk diagram with $\ell$ vertices for $\chi$ (say of type 0) inserted along the boundary and one vertex
for the NS-NS scalar $b$ in the interior, corresponding to the following amplitude
\begin{equation}
\label{abchi}
\big\langle \underbrace{V_{\chi_0}\cdots V_{\chi_0}}_{\ell}\,V_b\,\big\rangle~.
\end{equation}
By taking the $b$ vertex in the $(-1,-1)$ picture as in (\ref{verticesbs}), we saturate the
superghost number anomaly of the disk and thus we have to take all $\chi_0$ vertices in the
0 picture where they are simply
\begin{equation}
V^{(0)}_{\chi_0}=\frac{\ii}{2\pi}\,\chi_0\,\partial\bar{Z}(x)~.
\label{vertexchi0}
\end{equation}
Computing the CFT correlation functions among the vertex operators as discussed in
Ref.~\cite{Billo:2011uc}, and inserting the
topological normalization $\frac{\pi}{g_s}$ of the instantonic disk, one finds
that the amplitude (\ref{abchi}) for a given ordering of the open string vertices is
\begin{equation}
-\frac{2\ii\pi^2}{g_s}\,\frac{\bar p^\ell}{(2\pi)^\ell}
\,\tr_{k_0}\chi_0^\ell \,b
\int\frac{d^2w\prod_{i=1}^\ell dx_i}{dV_{\mathrm{CKG}}}\,
\,\frac{(w-\bar{w})^{\ell-2}}{\prod_{i=1}^\ell\lvert x_i-w\rvert^2}=
-\frac{\pi(\ii\bar p)^\ell}{g_s(\ell-1)!}
\,\tr_{k_0}\chi_0^\ell \,b~.
\label{abchi1}
\end{equation}
Summing over the $(\ell-1)!$ inequivalent orderings of the $\chi_0$ vertices (all of which give
the same contribution), multiplying by the symmetry factor $1/\ell\,!$ and summing over $\ell$
to account for all possible insertions, in the end
we find
\begin{equation}
 A_b=-\frac{\pi}{g_s}\,\sum_{\ell=0}^\infty\frac{(\ii\bar p)^\ell}{\ell\,!}
\,\tr_{k_0}\chi_0^\ell \,b = -\frac{\pi}{g_s}\,\tr_{k_0}\ee^{\ii\bar p \,\chi_0} \,b~.
\label{abchif}
\end{equation}

Let's finally compute the disk diagram with the insertion of a vertex for the R-R scalar $c$ in the interior and of $\ell$ $\chi_0$ vertices on the boundary, namely
\begin{equation}
\label{acchi}
\big\langle \underbrace{V_{\chi_0}\cdots V_{\chi_0}}_{\ell}\,V_c\,\big\rangle~.
\end{equation}
Putting the closed string vertex $V_c$ in the $(-\frac12,-\frac12)$ picture
as in (\ref{vc}), we saturate the superghost anomaly by taking $\ell-1$ vertices for $\chi_0$ in the 0 picture as in (\ref{vertexchi0}) and one vertex in the $-1$ picture where it reads
\begin{equation}
V^{(-1)}_{\chi_0}=\frac{\ii}{2\pi}\,\chi_0\,\bar{\psi}(x)\,\ee^{-\varphi(x)}~.
\end{equation}
Thus, the amplitude (\ref{acchi}) for a given ordering of the open string vertices becomes
\begin{equation}
-{2\pi^2}\,\frac{\bar p^\ell}{(2\pi)^\ell}
\,\tr_{k_0}\chi_0^\ell \,c
\int\frac{d^2w\prod_{i=1}^\ell dx_i}{dV_{\mathrm{CKG}}}\,
\,\frac{(w-\bar{w})^{\ell-2}}{\prod_{i=1}^\ell\lvert x_i-w\rvert^2}=
\frac{\ii\pi(\ii\bar p)^\ell}{(\ell-1)!}
\,\tr_{k_0}\chi_0^\ell \,c~.
\label{acchi1}
\end{equation}
Summing over all inequivalent orderings, taking into account the symmetry factors and summing
over $\ell$, we finally get
\begin{equation}
 A_c=\ii\pi\sum_{\ell=0}^\infty\frac{(\ii\bar p)^\ell}{\ell\,!}
\,\tr_{k_0}\chi_0^\ell \,c= \ii\pi\,\tr_{k_0}\ee^{\ii\bar p \,\chi_0} \,c~.
\label{acchif}
\end{equation}
Adding the two contributions (\ref{abchif}) and (\ref{acchif}), in the end we obtain the following
contribution to the instanton moduli action
\begin{equation}
 -\ii\pi\,\tr_{k_0}\ee^{\ii\bar p \,\chi_0}\,\big(c+\ii\frac{b}{g_s}\big) = -\ii\pi\,
\tr_{k_0}\ee^{\ii\bar p \,\chi_0} \,t
\end{equation}
as reported in the main text.

\section{The UV/IR relation for SU(2) and SU(3)}
\label{secn:appc}
Here we provide a derivation of the relations presented in Section \ref{secn:gaugegrav}
between the twisted field $t$ and the coupling constant $\tau$ for the SU(2) and SU(3) theories in the special vacuum, which is based on the use of the corresponding SW curves.

\paragraph{SU(2):}
We derive the relation (\ref{q0t1}) by exploiting the information
encoded in the SW curve for the SU(2) theory, namely
\begin{equation} 
y^2 = (z^2 + u_2)^2-g^2\,(z^2-m^2)^2~,
\end{equation}
and in its four roots
\begin{equation}
  z_{1,2} =\pm \sqrt{\frac{+u_2+g\,m^2}{g-1}}~,~~~z_{3,4} =\pm \sqrt{\frac{-u_2+g\,m^2}{g+1}}~.
\label{roots1}
\end{equation}
Notice that $z_1=z_3$ and $z_2=z_4$ in the classical limit $g\to 0$. As is well-known, the
complex structure parameter of the SW curve, which is identified with the effective
SU(2) coupling $\tau$, can be related to an anharmonic ratio
of the roots. In particular we have~\footnote{More often in the literature one finds the
following relation
\begin{equation*}
 \xi=\frac{\zeta}{\zeta-1}=\frac{(z_1-z_3)(z_2-z_4)}{(z_1-z_2)(z_3-z_4)}=
\frac{\theta_2^4(\tau)}{\theta_3^4(\tau)}~.
\end{equation*}
This is simply related to (\ref{zetatheta}) by the transformation $\tau\to\tau+1$.}
\begin{equation}
 \zeta = \frac{(z_1-z_3)(z_2-z_4)}{(z_1-z_4)(z_2-z_3)} = - 
\frac{\theta_2^4(\tau)}{\theta_4^4(\tau)}=
-16\,\frac{\eta^8(4\tau)}{\eta^8(\tau)}~.
\label{zetatheta}
\end{equation}
Substituting the values (\ref{roots1}) and taking into account (\ref{vv}), we have
\begin{equation}
 \zeta= \frac{{\mathbf v}-\frac{2q_0}{1+q_0}\,m^2-\sqrt{\big({\mathbf v}-\frac{2q_0}{1+q_0}\,m^2\big)^2
-\frac{4q_0}{(1+q_0)^2}\,\big({\mathbf v}-m^2\big)^2}}{{\mathbf v}-\frac{2q_0}{1+q_0}\,m^2+\sqrt{\big({\mathbf v}-\frac{2q_0}{1+q_0}\,m^2\big)^2-
\frac{4q_0}{(1+q_0)^2}\,\big({\mathbf v}-m^2\big)^2}}~.
\label{zetar}
\end{equation}
Comparing with the solution (\ref{tsugrapure}) and using the identification $z^2\leftrightarrow \mathbf{v}$, we conclude that $\zeta=\ee^{\ii\pi t}$ from which the relation (\ref{q0t1})
immediately follows.

\paragraph{SU(3):}
Let us now prove the relation (\ref{q0t3}) for the conformal SU(3) theory. 
According to Ref.~\cite{Minahan:1995er}, the SW curve for this
theory in the special vacuum can be put in the form
\begin{equation}
 y^2 = (z^3 + u_3)^2-f(\tau)\,z^6
\label{SWcurveMinah}
\end{equation}
where $f(\tau)$ is a function of the effective IR coupling which is given
in terms of the genus-2 Riemann $\Theta$-functions.
These are a generalization of the Jacobi $\theta$-functions on the torus and their 
definition is
\begin{equation}
\Theta\ch{\vec{a}}{\vec{b}}= \sum_{\vec{n}\in \mathbb Z^2}
\ee^{\,\frac{1}{2}(\vec{n}+\frac{\vec{a}}{2})^t\,\Omega\,(\vec{n}+\frac{\vec{a}}{2})
\,+\,\ii\pi\,(\vec{n}+\frac{\vec{a}}{2})\cdot\vec{b}}~,
 \label{Theta}
\end{equation}
where $\Omega$ is the period matrix of the genus-2 Riemann surface and the 2-vectors $\vec{a}$ and $\vec{b}$ are the so-called 
characteristics related to the periodicity properties along the homology cycles.
For the curve (\ref{SWcurveMinah}) describing the conformal SU(3) theory at the special vacuum
the period matrix takes the simple form (\ref{tautree3}), namely
\begin{equation}
 \Omega = \begin{pmatrix}
     \,2&~ 1 \,\\
     \,1& ~2 \,\\
     \end{pmatrix}\,\pi\ii\,\tau~,
\end{equation}
and consequently several identities among the $\Theta$-functions hold. In particular, 
one can check that
\begin{equation}
\begin{aligned}
\Theta\ch{0\,0}{1\,0} &= \Theta\ch{0\,0}{0\,1} = \Theta\ch{0\,0}{1\,1}~\equiv~\vartheta_1(\tau)~,\\
\Theta\ch{1\,0}{0\,0} &= \Theta\ch{0\,1}{0\,0} = \Theta\ch{1\,1}{0\,0}~\equiv~\vartheta_2(\tau)~,\\
\Theta\ch{0\,1}{1\,0} &= \Theta\ch{1\,0}{0\,1} = \Theta\ch{1\,1}{1\,1}~\equiv~\vartheta_3(\tau)~.\\
\end{aligned}
 \label{varthetas}
\end{equation}
Given these definitions, the function $f(\tau)$ in (\ref{SWcurveMinah}) turns out to be given by
\cite{Minahan:1995er}
\begin{equation}
\begin{aligned}
 f(\tau) &= -\frac{27\,\vartheta_1^4(\tau)\,\vartheta_2^4(\tau)\,\vartheta_3^4(\tau)}{
\big(\vartheta_2^2(\tau)+
\vartheta_3^2(\tau)\big)^2
\big(\vartheta_1^2(\tau)+\vartheta_3^2(\tau)\big)^2
\big(\vartheta_1^2(\tau)-\vartheta_2^2(\tau)\big)^2}~.
\end{aligned}
\label{ftau}
\end{equation}
In the limit $\mathrm{Im}\tau\to\infty$, we have the following expansion
\begin{equation}
 f(\tau)= -108\,\ee^{\ii\pi\tau}\Big(1+66\,\ee^{\ii\pi\tau}+3573\,\ee^{2\ii\pi\tau}+
175468\,\ee^{3\ii\pi\tau}+\cdots\Big)
~.
\label{fexp}
\end{equation}
On the other hand, as we have explained in Sections \ref{secn:chiral} and \ref{secn:gaugegrav}, 
the SW for the conformal SU(3) theory at the special vacuum can be written as
\begin{equation}
 y^2 = (z^3 +u_3)^2-\frac{4q_0}{(1+q_0)^2}\,z^6~.
\label{SWcurveus}
\end{equation}
Thus, comparing the two curves we deduce that
\begin{equation}
 \frac{4q_0}{(1+q_0)^2}= f(\tau)~.
\label{qf}
\end{equation}
Using the properties of the $\vartheta$-functions (\ref{varthetas}) and/or
the expansion (\ref{fexp}), after some algebra we obtain
\begin{equation}
 q_0=-27\,\frac{\eta^{12}(3\tau)}{\eta^{12}(\tau)}=-27 \,\big( \ee^{\ii\pi\tau} +12 \,\ee^{2\ii\pi\tau}+
90\, \ee^{3\ii\pi\tau}+ \cdots\big)
\end{equation}
which is the UV/IR relation for SU(3) reported in (\ref{q0t3}).

\providecommand{\href}[2]{#2}\begingroup\raggedright\endgroup


\end{document}